\documentclass[aps,prd,float,superscriptaddress,notitlepage,nofootinbib]{revtex4-1}

\usepackage[utf8]{inputenc}
\usepackage[english]{babel}
\usepackage{mathtools}
\usepackage{subdepth}
\usepackage{natbib}
\usepackage[bottom]{footmisc}

\begin{document}

\title{Geometrical Interpretation of Electromagnetism in a 5-Dimensional Manifold}

\author{TaeHun Kim}
\email[]{gimthcha@snu.ac.kr}
\affiliation{Department of Physics \& Astronomy, Seoul National University 1 Gwanak-ro, Gwanak-gu, Seoul, 08826, Republic of Korea}

\author{Hyunbyuk Kim}
\email[]{time@kaist.ac.kr}
\thanks{Author to whom any correspondence should be addressed.}
\affiliation{Department of Physics \& Earth Science, Korea Science Academy of KAIST, 105-47, Baegyanggwanmun-ro, Busanjin-gu, Busan, 47162, Republic of Korea}


\begin{abstract}
In this paper, Kaluza–Klein theory is revisited and its implications are elaborated. We show that electromagnetic 4-potential can be considered as a shearing-like deformation of a 5-dimensional (5D) manifold along the fifth (5th) axis. The charge-to-mass ratio has a physical meaning of the ratio between the movement along the direction of the 5th axis and the movement in the 4D space-time. In order to have a 5D matter which is consistent with the construction of the 5D manifold, a notion of particle-thread is suggested. Examinations on the compatibility of reference frames reveal a covariance breaking of the 5th dimension. The field equations which extend Einstein's field equations give the total energy-momentum tensor as a sum of that of matter, electromagnetic field, and the interaction between electric current and electromagnetic potential. Finally, the experimental implications are calculated for the weak potential case.
\end{abstract}


\maketitle

\allowdisplaybreaks

\section{Introduction} \label{section intro}
It has been nearly a hundred years since the attempt for a unified field theory was first made. Although nowadays mainstream concern about the unification is to obtain a quantum theory of gravity, the first approaches were to make a theory of electromagnetism in the manner of differential geometry of general relativity. The most representative work of this kind was done by Theodor Kaluza \cite{kaluza1921unitatsproblem} and Oskar Klein \cite{klein1926quantentheorie}. The following brief introduction of Kaluza–Klein theory is based on \cite{overduin1997kaluza, goenner2004history, van2002einstein}. In 1921, Kaluza tried to unify the electromagnetism and the gravitation in 5-dimensions. Kaluza constructed a 5D metric including a new scalar field and the whole metric was independent of the 5th dimension. He also assumed the source-free condition for 5-dimensions. The geodesic equation in 5-dimensions produced a term that could be thought as the Lorentz force. However, there was another term containing the scalar field and this term could exceed the Lorentz force term. This additional term was a critical defect of Kaluza's model.

In 1926, Klein proposed a modified version of Kaluza's 5D metric, which was
\begin{equation}
\tilde{g}_{\alpha \beta}=
\begin{pmatrix}
g_{ab}+\phi^2 A_a A_b & \phi^2 A_a \\
\phi^2 A_b & \phi^2
\end{pmatrix}, \label{KK}
\end{equation}
where \(\phi\) was a scalar field and the repeating boundary condition was imposed along the 5th dimension. This is what we now call as the Kaluza–Klein metric. Then, from Einstein's field equations for 5D vacuum, 4D Einstein's field equations with energy-momentum of electromagnetic field could be obtained when \(\phi\) was fixed to be a constant. The fixation of \(\phi\) could avoid the critical defect of Kaluza's model. Klein then compactified the 5th dimension in order to connect his model to quantum mechanics. However, it was discovered later that the fixation of \(\phi\) is consistent with the source-free condition only if \(F_{ab}F^{ab}=0\) \cite{thiry1948geometrie}.

In this paper, we consider a model to explain the electromagnetism as a geometrical structure in 5-dimensions. The present model does not compactify the 5th dimension, so it is closer to Kaluza's original idea. The suggested model obtains a metric which has an identical form of Klein's metric (or Kaluza–Klein metric) with a fixed \(\phi\) through a shearing-like deformation of a 5D manifold. Even though the metric has the same form, the construction of the metric for the 5D manifold and the 4D space-time (throughout the present paper, space-time will mean 4D space-time) is different from both Kaluza's original idea and Klein's version. The model focuses not on the induced matter from higher dimensional geometry but on the pre-existing matter. Because of this, the model does not inherit the source-free condition and gives a new form of 5D matter. 

The structure of the present paper is as follows. Motivations for introducing the 5th dimension and the method to get the 4D physics from the 5D one are discussed in section \ref{section design}. Then, the 5D manifold used throughout this paper is constructed and the interpretation of electromagnetic 4-potential is given at section \ref{section construction}. Here, we also suggest that 5D matter is composed of many 1D threads extending parallel to the 5th axis. Section \ref{section physical quantities} is about physical quantities in the deformed manifold and decides the metric of the 5D manifold. Discussions on the splitting of 5D field equations are also given here. Section \ref{section geodesic} shows that, with a help from the interpretation of the charge-to-mass ratio, the 4-acceleration due to the Lorentz force originates from the projection of a 5D geodesic onto the space-time. Section \ref{section covariance breaking} examines the compatibility of reference frames, and considers that the covariance of the 5th dimension is broken. Based on all the previous discussions about the 5th dimension, the 4D field equations which extend Einstein's field equations are finally suggested in section \ref{section field equations}, and this completes the discussions on the field equations. Here we show that the total energy-momentum tensor is a sum of energy-momenta of matter, electromagnetic field, and the interaction between electric current and electromagnetic potential. Section \ref{section unit system} deals with the unit system derived from the construction of the field equations. Section \ref{section experimental implications} discusses the experimental implications. Lastly, section \ref{section conclusion} gives the conclusion and suggests further researches.

Throughout the paper, indices for 4D variables are given in Roman alphabets (\(a\), \(b\), \(c\), ...), whereas indices for 5D variables are given in Greek alphabets (\(\alpha\), \(\beta\), \(\gamma\), ...). Tilde is used for physical quantities in the deformed 5D manifold and the deformed space-time. An upper bar is used when the indices alone cannot guarantee that the quantity is calculated in the 5D manifold. \(s\) is the 5D interval and \(\tau\) is the proper time in the space-time. Notations are summarized in appendix \ref{appendix notations}. We adopt the geometrized unit \(G=c=1\). Derivations of equations are also given in the appendices.

\section{Design of the model} \label{section design}
The model constructed in this paper is inspired from the characteristic of a static gravitational field. In a static gravitational field, the metric of space-time is time independent and 3D space at each moment becomes a 3D hypersurface of a constant time. While every object follows its geodesic in the space-time, its trajectory in a 3D space is curved and is not a 3D geodesic in space. The 3D trajectory is a projection of the 4D geodesic onto the 3D space and the curvature of 3D trajectory in space depends on the object's velocity in the Newtonian sense. Although the direction of velocity can be known in the 3D space by dealing with the 3D trajectory, the magnitude of velocity cannot be determined only from the 3D trajectory but should be found in the space-time. Interestingly, the magnitude of velocity indicates how much the 4D geodesic is parallel to the time axis such that the more parallel it is, the smaller is the speed. So, the curvature of 3D trajectory becomes smaller as the object's speed becomes higher, or as the 4D geodesic becomes less parallel to the time axis and more parallel to the 3D space. 

Analogous arguments can be applied to a model which describes electromagnetic field in a 5D manifold. Now electromagnetic field is a geometrical structure of the 5D manifold and every object follows 5D geodesic motion. Space-time is a 4D hypersurface with a constant 5th coordinate and the 5D metric is independent of the 5th coordinate. An object's trajectory on the space-time is a projection of its 5D geodesic onto the space-time and generally is not a 4D geodesic in space-time. The curvature of 4D trajectory is the 4-acceleration and this should be regarded as the effect of the Lorentz force. Meanwhile, it is known that the 4-acceleration due to the Lorentz force depends on the 4-velocity and the charge-to-mass ratio of an object. For a given electromagnetic field, it is experimentally verified that the direction of the 4-acceleration is decided by the 4-velocity and the magnitude of the 4-acceleration is directly proportional to the charge-to-mass ratio. 

Therefore, the influence of 4-velocity on the curvature of 4D trajectory in a 5D manifold corresponds to the influence of the direction of 3-velocity on the curvature of 3D trajectory in a static gravitational field. More importantly, the influence of the charge-to-mass ratio on the curvature of 4D trajectory in a 5D manifold corresponds to the influence of the magnitude of 3-velocity on the curvature of 3D trajectory in a static gravitational field (note that it is meaningless to consider the magnitude of 4-velocity). This comparison encourages the interpretation that an object's charge-to-mass ratio can be thought as a quantity indicating how much its 5D geodesic is parallel to the 5th axis or parallel to the space-time. Since higher charge-to-mass ratio gives higher 4-acceleration, objects with a higher charge-to-mass ratio should have 5D geodesics more parallel to the 5th axis, and neutral objects may have 5D geodesics parallel to the space-time. We propose more quantitatively that the charge-to-mass ratio is a ratio of the movement along the direction of the 5th axis to the movement on the space-time. This initial idea about charge-to-mass ratio in terms of a dimensionless particle will be supplemented and more completely interpreted using a 1D object in sections \ref{section construction} and \ref{section physical quantities}. 

Once the physics in the 5D manifold is fully described it should be reduced to that of the 4-dimensions in order to get the equations in terms of the \textit{observed} quantities in the space-time without considering the 5th dimension. This reduction will allow us to check the 5D physics with the experimentally verified 4D physics. The prescribed setup, especially the part that the space-time is the hypersurface with a constant 5th coordinate, makes the reduction process a simple projection. The coordinate system in the 5D manifold is capable of being constructed in such a way that its 4D part coincides with the coordinate of the space-time. From now on, that coordinate system is assumed. This coordinate construction of the 5D manifold gives the induced metric of the space-time to be identical to the 4D part of the 5D metric. Also, a trajectory of an object in the space-time can be simply obtained by taking the 4D part of the 5D geodesic due to this coordinate construction. 

Although it is common to use the 5th dimension to explain electromagnetism in a geometrical manner, the setup of the present model and the resultant reduction process along with the reduced 4D physics are distinct from the original Kaluza–Klein theory. Reduction in Kaluza–Klein theory can be found in \cite{overduin1997kaluza, duff1994kaluza, bailin1987kaluza}. In Kaluza–Klein theory, the way of embedding of space-time into 5D manifold is not specified but only the metric relation between 5D manifold and the space-time is given. Instead, in the present model, it is established that the space-time is a hypersurface with a constant 5th coordinate. This results in a different space-time metric from the Kaluza–Klein case. By showing that the Ricci scalars of the space-time for the two cases are different, it is clear that the space-time in the present model and the space-time in Kaluza–Klein theory are actually different, so the ways of embedding are also. Calculations of the Ricci scalar of the space-time is given in appendix \ref{appendix Ricci scalar}, after obtaining the final 5D metric via discussions in sections \ref{section construction} and \ref{section physical quantities}. The difference in embedding raises differences on the space-time metric (see section \ref{section physical quantities}), on the correction terms in the 4-acceleration of the space-time trajectory (see section \ref{section geodesic}), and even on the signature of the 5th dimension (see section \ref{section field equations}). The first two are physical and not mere mathematical differences as they can be tested by experiments (see section \ref{section experimental implications}).

\section{Construction of the model} \label{section construction}
Our model inherits the qualitative features of general relativity but in 5-dimensions, so that the 5D manifold is described by 5D metric and the metric is affected by 5D energy-momentum. We consider a 5D manifold which is constructed by \textit{dragging} the space-time to the direction of the 5th axis so that the 4D part of 5D metric is independent of the 5th coordinate. However, to validate the independence of 5D metric on the 5th dimension, the independence of matter on the 5th dimension is needed. Therefore, we consider that particles in space-time are also dragged together so that they become \textit{threads} parallel to the 5th axis. Given the notion of the `particle-thread', a particle in the space-time becomes the cross section of the corresponding particle-thread along the space-time. 5D matter composed of particle-threads ensures that the matter distribution is independent of the 5th coordinate, and hence the independence of whole 5D metric on the 5th coordinate due to the translational symmetry of matter. 

Actually, the notion of particle-thread is not additionally introduced but follows from the dragging construction. Considering that the 5D manifold is constructed by continuously dragging the space-time, a 5D point particle distribution is unnatural since placing discrete 5D matters will invalidate the construction process of dragging. Discrete placing means that there should be some unknown external factors affecting the process of dragging for deciding on which 4D hypersurface will the 5D particle be placed. Yet, no such external factors are allowable for our construction of the 5D manifold. 

In addition, the notion of particle-thread helps to define the distance between particles. If a point particle does not extend along the 5th dimension, the proposed interpretation of the charge-to-mass ratio implies that the 5D distance between two particles with different charge-to-mass ratios will diverge as time goes on, even if they are stationary in the 3D space. On the other hand, the distance between two particle-threads can be defined naturally as the distance between their cross sections on the space-time and then it is reduced to the well-established distance between the corresponding particles in the 3D space. 

Besides these helping features, it is possible to imagine a mechanical wave propagating along a particle-thread. A wave will create a perturbation on the energy-momentum of the particle-thread but the change of the energy-momentum along the 5th dimension should be negligible in large scale. Although the possibility of a mechanical wave does not seem to raise any immediate issues, it should be examined thoroughly to see whether there is any contradiction with the observations. Also, the physical meanings of quantities associated with the wave, such as its wavelength, amplitude, and propagation speed, need to be investigated. The present paper will not pursue the issue related to the wave along a particle-thread further and deals only with the zero mode of the wave. 

The design of the 5D model in section \ref{section design} is now to be revised to a particle-thread version. Fortunately, both particle-thread and 5D metric are independent of the 5th dimension, so any pair of geodesics which have the same tangent vectors at each of their starting points with common space-time coordinates but different 5th coordinate are exactly identical. In other words, every portion of particle-thread moves identically and shows exactly the same behavior as a geodesic motion of a test particle in the 5D manifold. This guarantees that a particle-thread which is following its 5D geodesic does not entangle with another particle-threads along its extension unless there exists some pre-given internal motion of the particle-thread. The projection of 5D geodesic on the space-time will be the trajectory of the cross section of the particle-thread along the space-time. However, the charge-to-mass ratio should no longer be thought of as the ratio of a particle's movement along the 5th axis to its movement on the space-time, but as the \textit{flow speed} of a particle-thread along its extension. Then a particle-thread corresponding to a charged particle flows along its extension and a particle-thread corresponding to a neutral particle does not flow but only moves parallel to the space-time.

Now we start to construct the 5D manifold by considering first a situation where there exist only neutral objects. Starting from a space-time with its metric tensor \(g_{ab}\), a 5D manifold is constructed by \textit{dragging} the space-time into the 5th direction. This dragging construction makes every hypersurface of constant \(x^4\) to be identical to the space-time, so the 5D manifold will have a metric tensor 
\begin{equation}
g_{\alpha \beta}=
\begin{pmatrix}
g_{ab} & * \\
* & 1
\end{pmatrix},
\end{equation}
with a proper coordinate along the 5th dimension and every component of the metric is independent of the 5th dimension. We will take the signature of the 5th dimension to be positive. The \(4a\)-part of the metric decides the orthogonality of the space-time and the 5th direction. However, for the case that every object is neutral, all the particle-threads are not flowing so the 5D matter distribution will have a 5th axis-reversal symmetry. Therefore, the whole 5D manifold also should have the 5th axis-reversal symmetry and this gives 
\begin{equation}
g_{\alpha \beta}=
\begin{pmatrix}
g_{ab} & 0 \\
0 & 1
\end{pmatrix}.
\label{undeformed metric}
\end{equation}
Let us take this to be an electromagnetic field-free manifold. 

Now we consider a general situation where there exist charged objects too. In this case, the 5th axis-reversal symmetry is broken so it is possible to have non-vanishing \(4a\)-part. In our model, a flowing particle-thread trails its neighboring part of the 5D manifold so that \textit{shearing} occurs on the manifold in the direction of the flow of particle-thread. To represent the shearing quantitatively, we introduce an operation on a manifold which will be called deformation. 

A deformation is an operation that keeps the coordinate values of every point of the manifold while changing the manifold. This means that \(d\tilde{x}^\alpha = dx^\alpha\) for any infinitesimal displacement since the coordinate does not change, but its norm can differ since the manifold and its metric changes. Even though the effect of a deformation is completely different from that of a coordinate transformation, interestingly, they can give a similar-looking change of metric. Keeping this in mind, a deformation acting on a manifold gives a deformed metric by
\begin{equation}
\tilde{g}_{\alpha \beta}=D_\alpha^\gamma D_\beta^\delta g_{\gamma \delta}, \label{general deformation}
\end{equation}
where \(D_{\alpha}^{\gamma}\) performs the deformation. It is important to note that the Riemann tensor, the Ricci tensor, and the Ricci scalar of \(\tilde{g}_{\alpha \beta}\) cannot be computed by treating \(D_{\alpha}^{\gamma}\) as a coordinate transformation on \(g_{\gamma \delta}\) and applying the tensor transformation rule on those tensors since, again, the deformation is an operation on the manifold itself. More discussions about the deformation are given in appendix \ref{appendix deformation}. 

For our case, to find a shearing deformation, we first look at a coordinate transformation given by 
\begin{subequations}
\begin{eqnarray}
{x'}^{a}&=&x^a, \\
{x'}^{4}&=&-C_a x^a +x^4, 
\end{eqnarray}
\end{subequations}
where \(C_a=C_a(x^b)\). Then, the equations for the coordinate differentials become
\begin{subequations}
\begin{eqnarray}
{dx'}^{a}&=&dx^a, \\
{dx'}^{4}&=&-\left(C_a+C_{b,a}x^b\right) dx^a +dx^4,
\label{dx'4 coordinate transformation}
\end{eqnarray}
\end{subequations}
and we rename the whole parenthesis part in equation (\ref{dx'4 coordinate transformation}) to be \(C_a\) for the notational convenience. Then, the equations for the coordinate differentials are re-written as
\begin{subequations}
\begin{eqnarray}
{dx'}^{a}&=&dx^a, \\
{dx'}^{4}&=&-C_a dx^a +dx^4.
\end{eqnarray}
\end{subequations}
This transformation represents a shearing along the 5th dimension since the 5th axis remains the same but the other four axes become inclined while keeping the values of the first four coordinates. After the above transformation, the metric of equation (\ref{undeformed metric}) is now expressed in the primed coordinate,
\begin{equation}
{g'}_{\alpha \beta}=
\begin{pmatrix}
g_{ab}+C_a C_b & C_a \\
C_b & 1
\end{pmatrix}.
\label{metric from coordinate transform}
\end{equation}
This is the same form of Kaluza–Klein metric with \(\phi^2 = 1\). However, it is not true that all Kaluza–Klein metrics with \(\phi^2 = 1\) can be thought as a product of a coordinate transformation from the metric of the electromagnetic field-free manifold since the commutativity of the second partial derivatives requires \(C_{a,b} = C_{b,a}\).

Because of this, we consider a deformation which resembles the resulting equation of a coordinate transformation, equation (\ref{metric from coordinate transform}), but now the commutativity of the second partial derivatives is no longer of concern. To emphasize this, instead of \(C_a\), we introduce \(B_a\) which does the role of deformation. Due to the introduction of particle-thread, the deformation is independent of the 5th coordinate and hence \(B_a\) is a function of the space-time coordinates alone. Then the deformation given by
\begin{equation}
D_\alpha^\beta=
\begin{cases}
1, & \text{if} \ \alpha=\beta \\
B_a & \text{if} \ \alpha=a \ \text{and} \ \beta=4 \\
0, & \text{otherwise}
\end{cases}
\end{equation}
yields, using equations (\ref{undeformed metric}) and (\ref{general deformation}), the deformed 5D manifold with metric
\begin{equation}
\tilde{g}_{\alpha \beta}=
\begin{pmatrix}
g_{ab}+B_a B_b & B_a \\
B_b & 1
\end{pmatrix},
\label{deformed metric}
\end{equation}
where every component is independent of the 5th coordinate and \(B_{a,b}\) is no longer required to be the same with \(B_{b,a}\). A comparison between $\tilde{g}_{\alpha \beta}$ and ${g'}_{\alpha \beta}$ presents that this deformation affects the 5D manifold in the sense of shearing. It should be noted that since the deformation is not a mere mathematical handling but a physical phenomenon, the \(B_a\) in equation (\ref{deformed metric}) is the order parameter deciding how much the distance structure of the manifold changes. 

We will interpret \(B_a\), the quantity indicating the shearing-like deformation, as the electromagnetic 4-potential on each point of the deformed 5D manifold. This is justified by the fact that the 4-acceleration of a projection of 5D geodesic of that metric on the space-time gives the Lorentz force term as the leading term (the calculation will be given in section \ref{section geodesic}). Then the metric \(\tilde{g}_{\alpha \beta}\) of the deformed 5D manifold becomes Kaluza–Klein metric with \(\phi^2 = 1\). Before deriving more equations using this metric of deformed manifold, the electromagnetic 4-potential should be defined clearly. This discussion will be continued in section \ref{section physical quantities}.

Is the 5th dimension space-like or time-like? It depends on the choice of the signature of the space-time metric. In section \ref{section field equations}, the field equations reveal that the 5th dimension is time-like by showing that the 4D signature should be ($+$ $-$ $-$ $-$). But then the ultrahyperbolicity problem arises. As Tegmark pointed out \cite{tegmark1997dimensionality}, the partial differential equations in two or more time dimensions become ultrahyperbolic and the world becomes unpredictable. Remarkably, the particle-thread of the present model hinders this problem. The particle-thread guarantees both metric and energy-momentum to be independent of the 5th dimension, and thanks to this, every partial derivative with respect to the 5th coordinate vanishes. Then, the 5D partial differential equations can be recast into the hyperbolic 4D partial differential equations of the space-time. In this respect, one can say that the particle-thread does the role of keeping the hyperbolicity of the 5D manifold.

\section{Physical quantities in the deformed manifold} \label{section physical quantities}
Since a deformation acting on a manifold affects the manifold itself, all the observable quantities in the presence of non-zero electromagnetic potential should be defined in the deformed manifold. In a sense, an undeformed manifold performs only the intermediate role. The fact, which will be proven in section \ref{section geodesic}, that the Lorentz force does appear in the deformed manifold supports this. Therefore, all the observable quantities in the space-time including the Lorentz force and the 4-velocities will be defined in the deformed space-time. In other words, the deformed space-time is the right stage where the 4D physics is described.

We also define the charge-to-mass ratio of an object in the deformed manifold. It is interpreted as the flow speed of the corresponding particle-thread along its extension, and its every portion follows an identical 5D geodesic. For this geodesic, the charge-to-mass ratio has a meaning of the ratio of movement along the direction of the 5th axis to the movement on the space-time. Therefore, an object's charge-to-mass ratio can be expressed by
\begin{equation}
\frac{q}{m}=\frac{dx^{4}}{d\tilde{\tau}}, \label{equation charge-to-mass ratio}
\end{equation}
for the geodesic that every portion of the corresponding particle-thread follows. This allows us to picture the charge-to-mass ratio as a slope of the geodesic in 5-dimensions.

For a particle-thread, its cross section along the space-time is a particle in the space-time, and mass is a character of that cross section. Therefore, the electric charge \(q=m\times (dx^{4}/d\tilde{\tau})\) becomes a quantity describing the mass flow rate of a particle-thread. From this, the picture of the model can be drawn such that the mass flow of a particle-thread along the 5th direction in the 5D world is conceived as an electric charge in a 4D hypersurface. So, in the present model, there is no such thing as a 3D brane in which the matters are confined, nor is there any division between the brane and the bulk.

As discussed in section \ref{section construction}, 5D matter is composed of particle-threads. This suggests that two physical quantities are required to describe 5D matter. One is a 4D energy-momentum tensor which corresponds to the distribution and movement of cross sections of particle-threads on the space-time. The other is a 4-current density which describes the flux of the particle-threads. These two are enough to describe 5D matter distribution considering that the only appropriate form of 5D matter is a particle-thread. Actually, using 4D quantities only is better than newly introducing a 5D energy-momentum tensor because the 5D volume density of a particle-thread blows up. Note that the 4D surface density of the 5D matter along the space-time is the mass distribution which is finite and non-zero.

For a given matter distribution, the two quantities, 4D energy-momentum tensor and 4-current density, can be obtained by conventional means but now in the deformed space-time. This disjunction of 5D matter quantities and the covariance breaking of the 5th dimension, which is dealt with in section \ref{section covariance breaking}, causes the field equations relating the 5D metric and the 5D matter to be described by two split equations, one using only the 4-current density and the other using both. 

Before discussing more about the field equations, the metric of the 5D manifold should be clarified first. As already mentioned in section \ref{section construction}, the electromagnetic 4-potential does the role of the deformation factor \(B_{a}\) of the deformed 5D manifold with respect to the undeformed 5D manifold. This deformation is caused by flowing particle-threads trailing the surrounding 5D manifold, and the flux of the particle-threads is the electric 4-current in the \textit{deformed space-time}, which is an observable quantity. Therefore, the electromagnetic 4-potential will be defined in the deformed space-time as well, unlike Kaluza–Klein theory. Then, by replacing \(B_a\) in equation (\ref{deformed metric}) to \(\tilde{A}_a\), emphasizing that it is the electromagnetic 4-potential defined in the deformed space-time, the metric tensor of the deformed 5D manifold can be written as
\begin{equation}
\tilde{g}_{\alpha \beta}=
\begin{pmatrix}
g_{ab}+\tilde{A}_a \tilde{A}_b & \tilde{A}_a \\
\tilde{A}_b & 1
\end{pmatrix}
=
\begin{pmatrix}
\tilde{g}_{ab} & \tilde{A}_a \\
\tilde{A}_b & 1
\end{pmatrix},
\label{deformed 5D metric}
\end{equation}
where the second equality comes from the discussion on the induced metric tensor of the deformed space-time in section \ref{section design}. Here, every component is independent of the 5th coordinate. Again, as noted in section \ref{section construction}, \(\tilde{A}_a\) does the role of an order parameter indicating the change of 5D manifold.

Now we resume the discussion on the field equation that employs only the 4-current density. It will be shown in section \ref{section geodesic} that when the shearing-like deformation factor is interpreted as the electromagnetic 4-potential, the leading term of 4-acceleration from the projection of 5D geodesic onto the space-time becomes the Lorentz force. From the well established facts that the Lorentz force depends on the electromagnetic fields and these fields follow Maxwell's equations, the equations determining \(\tilde{A}_{a}\) will be Maxwell's equations. However, unlike the conventional Maxwell's theory, the issue of gauge condition arises for \(\tilde{A}_{a}\) in this 5-dimensional model because now the physical quantity is not the electromagnetic field but the electromagnetic 4-potential itself. Thus, the gauge of the electromagnetic 4-potential must be fixed. The most natural way of gauge fixing is to take it to be the \textit{retarded potential in the deformed space-time} to satisfy the causality in the space-time. In short, the field equations for \(\tilde{A}_{a}\), or \(\tilde{g}_{4a}\), will be \textit{Maxwell's equations for electromagnetic 4-potential in Lorenz gauge in the deformed space-time} and the physically valid solution of these equations is a retarded potential due to the 4-current density \(\tilde{J}^b\) in the deformed space-time. The general form of this solution is
\begin{equation}
\tilde{A}_a(x) = \int \tilde{G}^{\rm{\langle ret \rangle }}_{ab}(x, x') \tilde{J}^b(x') \sqrt{-\tilde{g}(x')} d^4x', 
\label{retarded potential}
\end{equation}
where the integral is taken over the deformed space-time and \(\tilde{G}^{\rm{\langle ret \rangle }}_{ab}(x, x')\) is the Green's function for retarded potential. \(\tilde{G}^{\rm{\langle ret \rangle }}_{ab}(x, x')\) and \(\tilde{g}(x')\) are functions of the metric of the deformed space-time and, as a consequence, functions of the 4-potential itself. To obtain a complete equation for the 4-potential, the Green's function should be expressed in terms of the quantities associated with the deformed space-time such as the metric, the Riemann tensor, the Ricci tensor, and the Ricci scalar. This is discussed in \cite{poisson2011motion}. Instead of pursuing the exact Green's function of the deformed space-time in this paper, we will consider a weak potential approximation in section \ref{section experimental implications}. Until now, we have dealt with the field equations for \(\tilde{g}_{4a}( = \tilde{A}_a\)), so the only part remaining to be determined in 5D metric is \(\tilde{g}_{ab} = g_{ab}+\tilde{A}_{a} \tilde{A}_{b}\). The field equations for this part will be suggested in section \ref{section field equations} after discussing the covariance breaking of the 5th dimension in section \ref{section covariance breaking}. 

At this moment, the features of metric and the metric construction separate the present model from Kaluza–Klein theory. In Kaluza–Klein theory, 5D metric is given as an `ansatz', but in the present model, 5D metric is obtained by dragging the space-time and applying the shearing-like deformation to the 5D manifold. Also, the source of the deformation is said to be particle-threads of charged objects, giving us a picture that flowing particle-threads trail the surrounding 5D manifold. Furthermore, although the two expressions of the 5D metric in equation (\ref{deformed 5D metric}) look similar to Kaluza–Klein metric and Kaluza's original metric, respectively, the decomposed parts of the 5D metric are different. The former expression in equation (\ref{deformed 5D metric}) has the electromagnetic 4-potential defined on the deformed space-time unlike Kaluza–Klein metric that employs electromagnetic 4-potential defined on the undeformed space-time. Also, the space-time metric for the latter expression in equation (\ref{deformed 5D metric}) is \(\tilde{g}_{ab}\) instead of \(g_{ab}\) used in Kaluza's original metric. All these differences come from the embedding of the space-time. Since all the physical quantities are defined in the deformed manifold, we will use the latter expression of equation (\ref{deformed 5D metric}) except the cases where it is needed to distinguish the effect of electromagnetic field and the effect of matter without electromagnetic field separately.

By taking inverse of \(\tilde{g}_{\alpha \beta}\), the inverse metric tensor of the deformed 5D manifold becomes
\begin{equation}
\tilde{g}^{\alpha \beta}=
\begin{pmatrix}
\tilde{g}^{ab} +\tilde{k}\tilde{A}^a\tilde{A}^b &  -\tilde{k}\tilde{A}^a \\
-\tilde{k}\tilde{A}^b & \tilde{k}
\end{pmatrix},  \quad \text{where} \quad 
\tilde{k}=\frac{1}{1-\tilde{A}_c \tilde{A}^c}. 
\label{inverse deformed 5D metric}
\end{equation}
Equations (\ref{deformed 5D metric}) and (\ref{inverse deformed 5D metric}) include the relation between the metric and the inverse metric of the deformed space-time. For the 4D physics, index raising and lowering of 4D tensors should be done in the space-time without referencing the 5D manifold. 

This way of defining the electromagnetic 4-potential prevents loopholes in the model building by making all the observed quantities defined on the deformed space-time. But, at the same time, this makes the calculation of electromagnetic 4-potential with a given 4-current density difficult. To calculate \(\tilde{A}_a\), the retardation must be decided by \(\tilde{g}_{ab}\) which is the function of \(\tilde{A}_a\) itself. Therefore, to know \(\tilde{A}_a\) for a given 4-current density distribution, \(\tilde{g}_{ab}\) and \(\tilde{A}_a\) should be determined simultaneously. Fortunately, there is a more convenient way if the 4-current density is weak enough to make \(\tilde{A}_a\tilde{A}_b\) sufficiently small compared to \(g_{ab}\). In this case, it will be acceptable to calculate \(\tilde{A}_a\) in the undeformed space-time as the zeroth order of approximation.

\section{Projection of 5D geodesic} \label{section geodesic}
Now we show that the suggested geometrical interpretation of the charge-to-mass ratio and the electromagnetic 4-potential gives the right 4-acceleration which can be seen as the Lorentz force for the projection of 5D geodesic onto the space-time. In the deformed 5D manifold we shall give the 5D geodesic equation in terms of the 4D proper time (\(\tilde{\tau}\)) since the Lorentz force will be expressed with a 4-velocity. The 4D proper time is not an affine parameter, so
\begin{eqnarray}
\frac{d^2x^{\alpha}}{d\tilde{\tau}^2}+\tilde{\Gamma}^{\alpha}_{\phantom{\alpha} \beta \gamma}\frac{dx^{\beta}}{d\tilde{\tau}}\frac{dx^{\gamma}}{d\tilde{\tau}}&=&f(\tilde{\tau})\frac{dx^{\alpha}}{d\tilde{\tau}}. \label{equation geodesic tau}
\end{eqnarray}
A lengthy calculation (appendix \ref{appendix f tau}) gives
\begin{equation}
f(\tilde{\tau})=-\tilde{g}_{4a}\tilde{\Gamma}^{4}_{\phantom{4}\beta\gamma}\frac{dx^{\beta}}{d\tilde{\tau}}\frac{dx^{\gamma}}{d\tilde{\tau}}\frac{dx^{a}}{d\tilde{\tau}}. \label{equation f tau}
\end{equation}
The 4D part of the geodesic equation in the deformed 5D manifold becomes
\begin{widetext}
\begin{equation}
\frac{d^2x^{a}}{d\tilde{\tau}^2}
+\bar{\tilde{\Gamma}}^{a}_{\phantom{a}bc}\frac{dx^{b}}{d\tilde{\tau}}\frac{dx^{c}}{d\tilde{\tau}}
=-2\tilde{\Gamma}^{a}_{\phantom{a}4b}\frac{dx^{4}}{d\tilde{\tau}}\frac{dx^{b}}{d\tilde{\tau}}
-\tilde{\Gamma}^{a}_{\phantom{a}44}\frac{dx^{4}}{d\tilde{\tau}}\frac{dx^{4}}{d\tilde{\tau}}
-\left(\tilde{g}_{4d}\tilde{\Gamma}^{4}_{\phantom{4}\beta\gamma}\frac{dx^{\beta}}{d\tilde{\tau}}\frac{dx^{\gamma}}{d\tilde{\tau}}\frac{dx^{d}}{d\tilde{\tau}}\right)\frac{dx^{a}}{d\tilde{\tau}}.
\label{equation geodesic projection initial}
\end{equation}
\end{widetext}
This is the equation of the projected 5D geodesic onto the space-time. The Christoffel symbols are given in appendix \ref{appendix Christoffel deformed}. Then, rearrangement of terms finalizes the equation of the 4-acceleration in the deformed space-time (appendix \ref{appendix geodesic projection})
\begin{eqnarray}
\tilde{a}^{a} = \tilde{L}^{a} +\frac{1}{1-\tilde{A}_e \tilde{A}^e}\tilde{L}^{c}\tilde{A}_{c}\tilde{D}^{a}+\frac{1}{1-\tilde{A}_e\tilde{A}^e}(\tilde{\nabla}_b \tilde{A}_c)\frac{dx^b}{d\tilde{\tau}}\frac{dx^c}{d\tilde{\tau}}\tilde{D}^a, \label{equation geodesic projection final}
\end{eqnarray}
where 
\begin{eqnarray}
\tilde{a}^a = \frac{d^2x^{a}}{d\tilde{\tau}^2}
+{\tilde{\Gamma}}^{a}_{\phantom{a}bc}\frac{dx^{b}}{d\tilde{\tau}}\frac{dx^{c}}{d\tilde{\tau}}, \quad
\tilde{L}^a = \frac{dx^{4}}{d\tilde{\tau}}\tilde{F}^{a}_{\phantom{a}b}\frac{dx^{b}}{d\tilde{\tau}}, \quad \text{and} \quad
\tilde{D}^a = \tilde{A}^a - \tilde{A}_d \frac{dx^d}{d\tilde{\tau}} \frac{dx^a}{d\tilde{\tau}}.
\label{aLD in 4-acc}
\end{eqnarray}
The term \(\tilde{L}^a\) is the Lorentz force term. Equation (\ref{equation geodesic projection final}) shows the effect of deformation clearly. It can be easily checked that the 4-acceleration (and hence the Lorentz force) vanishes when there is no deformation because the electromagnetic 4-potential becomes zero in that case.

The equation also shows that, in addition to the Lorentz force term, there are corrections for the 4-acceleration in the present model. For Kaluza–Klein theory, there are no such additional correction terms because the proper time is defined on \(g_{ab}\) instead of \(\tilde{g}_{ab}\) \cite{kerner2000geodesic}. These correction terms are experimentally verifiable and can be used to distinguish the present model from Kaluza–Klein theory. We will calculate the explicit components of those terms in section \ref{section experimental implications}.

\section{Covariance breaking of the 5th dimension} \label{section covariance breaking}
The presented interpretation of the charge-to-mass ratio implies that the neutral particle-threads move only along the space-time while charged particle-threads flow along the direction of the 5th axis also. But then the motion will be relative between the particle-threads, so should a charged one observe itself as a neutral one and an originally neutral one as a charged one? An examination on the compatibility of reference frames gives the answer to this question. 

All of the electromagnetic phenomena have been described and successfully explained by an observer who observes nearly all of the macroscopic objects in the world to be neutral. We will call this observer a \textit{natural observer} and its frame as a natural frame. The suggested model in this paper produces a verified motion of an object at least to the leading term, the Lorentz force term, when the metric of the 5D manifold is constructed by a natural observer. We can examine whether a given frame is compatible with a natural frame by checking that the metric of the 5D manifold constructed by the corresponding observer gives the same 5D manifold constructed by a natural observer and also gives a correct motion for an object in a natural frame.

We denote a given frame with an acute accent and a natural frame without any special mark. Then a generic coordinate transformation between these two can be expressed as
\begin{subequations}
\begin{eqnarray}
{d\acute{x}}^{a}&=&\Lambda^a_b dx^b + M^a dx^4, \\
{d\acute{x}}^{4}&=&N_a dx^a + k \ dx^4.
\end{eqnarray}
\end{subequations}
Here, only the coefficients \(M^a\) and \(N_a\) are important for the present purpose, so we will take \(\Lambda^a_b = \delta^a_b\) and \(k = 1\). Then the equations become
\begin{subequations}
\begin{eqnarray}
{d\acute{x}}^{a}&=&dx^a + M^a dx^4, \\
{d\acute{x}}^{4}&=&N_a dx^a + dx^4.
\end{eqnarray}
\end{subequations}

Now we show that \(M^a\) should vanish in order to have a proper metric of the 5D manifold. The metric of the 5D manifold constructed by the given observer does not change with respect to \(\acute{x}^4\) so that \(\acute{g}_{\alpha \beta,4}=0\). Then the derivative of metric in a natural frame with respect to \(x^4\) is given by 
\begin{equation}
g_{\alpha \beta,4} = M^a \frac{\partial}{\partial \acute{x}^a}\left(\frac{\partial \acute{x}^\gamma}{\partial x^\alpha}\frac{\partial \acute{x}^\delta}{\partial x^\beta}\right) \acute{g}_{\gamma \delta} + M^a \frac{\partial \acute{x}^\gamma}{\partial x^\alpha}\frac{\partial \acute{x}^\delta}{\partial x^\beta} \acute{g}_{\gamma \delta,a} + \frac{\partial}{\partial \acute{x}^4}\left(\frac{\partial \acute{x}^\gamma}{\partial x^\alpha}\frac{\partial \acute{x}^\delta}{\partial x^\beta}\right) \acute{g}_{\gamma \delta}.
\end{equation}
Since we do not have any general relation between ${g}_{\alpha \beta}$ and ${g}_{\alpha \beta,a}$ (or between $\acute{g}_{\gamma \delta}$ and $\acute{g}_{\gamma \delta,a}$), the second term should vanish by itself in order to have \(g_{\alpha \beta,4}=0\). Then, to make \(M^a \acute{g}_{\gamma \delta,a}\) to be zero for arbitrary cases, \(M^a\) should be zero. Therefore, we conclude \(M^a\) must vanish for a compatible frame. 

Next we show that \(N_a\) also should vanish in order to have a proper interaction between particles in the space-time observed by a natural observer. For a given \(N_a\), we can find an object for which the motion of every portion of the corresponding particle-thread satisfies \(dx^4=-N_adx^a\neq0\) in a natural frame. Then let us assume that there are two identical objects satisfying this equation. We can put all other objects infinitely far away from these two objects in the natural frame and this allows us to deal with only the two objects in the given frame too. While these two objects are charged identically in the natural frame they are neutral in the given frame. If the given observer constructs the metric of the 5D manifold, there is no charged object in its frame within finite distance so \(\acute{A}_a = 0\) in any finite region including the two objects. So the two objects will not have 4-acceleration in the given frame. Now if their motion is observed by the natural observer, there might be a 4-acceleration which arises from the coordinate transformation including \(N_a\). However, this cannot give a correct 4-acceleration in the natural frame since this 4-acceleration now depends only on the details of \(N_a\) which has nothing to do with the physical condition such as the distance between the two objects. Thus we conclude \(N_a\) also should vanish for a compatible frame. 

To summarize, we cannot construct a compatible frame through a coordinate transformation from a natural frame if there exists any non-zero cross term between the space-time and the 5th dimension. That is, the covariance of the 5th dimension is broken. This can be understood in two aspects. One is \(M^a=0\) which means that every observer should agree on the 5th axis and the other is \(N_a=0\) which indicates that every observer shares a \textit{common neutral plane}, the 4D hypersurface in the 5D manifold, coinciding with the space-time. The covariance breaking might be associated with the dragging construction but the exact mechanism remains to be discussed elsewhere. 

The covariance is broken only for the 5th dimension and it results in the common 5th axis and common neutral plane. Since any motion along other than the 5th dimension is parallel to the neutral plane, the covariance of the other 4-dimensions is unaffected. Because of this, for a general coordinate transformation between the compatible observers, the 4D part of the 5D tensor behaves as a 4D tensor:
\begin{equation}
\acute{\bar{X}}_{a_1 \dots a_i}^{b_1 \dots b_j} = \frac{\partial x^{\gamma_1}}{\partial \acute{x}^{a_1}} \dots \frac{\partial x^{\gamma_i}}{\partial \acute{x}^{a_i}}\frac{\partial \acute{x}^{b_1}}{\partial x^{\delta_1}} \dots \frac{\partial \acute{x}^{b_j}}{\partial x^{\delta_j}} X_{\gamma_1 \dots \gamma_i}^{\delta_1 \dots \delta_j} = \frac{\partial x^{c_1}}{\partial \acute{x}^{a_1}} \dots \frac{\partial x^{c_i}}{\partial \acute{x}^{a_i}}\frac{\partial \acute{x}^{b_1}}{\partial x^{d_1}} \dots \frac{\partial \acute{x}^{b_j}}{\partial x^{d_j}} \bar{X}_{c_1 \dots c_i}^{d_1 \dots d_j}.
\end{equation}

\section{Field equations and the Energy-Momentum Tensor} \label{section field equations}
In this section we suggest the field equations for  \(\tilde{g}_{ab}\) which are needed to determine the deformed manifold completely in conjunction with \(\tilde{A}_a\) discussed in section \ref{section physical quantities}. The 4D part of the Einstein tensor of the deformed 5D manifold becomes (appendix \ref{appendix Einstein tensor})
\begin{eqnarray}
\bar{\tilde{G}}_{ab}=G_{ab}+\frac{1}{2}\left(\frac{1}{4}\tilde{g}_{ab}\tilde{F}_{cd}\tilde{F}^{cd}-\tilde{F}^{\phantom{a}c}_{a}\tilde{F}_{bc}\right)-\frac{1}{2}(\tilde{A}_a \tilde{\nabla}_c \tilde{F}^{c}_{\phantom{c}b} + \tilde{A}_b \tilde{\nabla}_c \tilde{F}^{c}_{\phantom{c}a}), \label{equation Einstein tensor}
\end{eqnarray}
up to the second power of the electromagnetic 4-potential, the Riemann tensor, the Ricci tensor, and the Ricci scalar of the undeformed space-time (the latter three will be called the curvature quantities of the undeformed space-time). For the 4D signature of ($+$ $-$ $-$ $-$), the above equation can be written as  
\begin{equation}
\bar{\tilde{G}}_{ab}=G_{ab}+\frac{\mu_0}{2}\tilde{T}_{ab}^{\rm{\langle EM \rangle }}-\frac{\mu_0}{2}(\tilde{A}_a \tilde{J}_b + \tilde{J}_a \tilde{A}_b), \label{equation Einstein tensor simplified}
\end{equation}
where the superscript $\langle$EM$\rangle$ stands for electromagnetic field. Now we choose the permeability of vacuum to be \(\mu_0 = 16\pi \). Also, by taking \(\bar{\hat{G}}_{ab}=8\pi \hat{T}_{ab}\) for the field equations of any manifold (i.e. deformed and undeformed manifold, this is emphasized with the hat mark), the field equations of the undeformed manifold become  \(G_{ab} = \bar{G}_{ab} = 8\pi T_{ab}^{\rm{\langle M \rangle }}\) (where the superscript $\langle$M$\rangle$ stands for matter). Finally, the field equations for the deformed manifold crystallize into
\begin{equation}
\bar{\tilde{G}}_{ab}=8\pi \left[T_{ab}^{\rm{\langle M \rangle }}+\tilde{T}_{ab}^{\rm{\langle EM \rangle }}-(\tilde{A}_a \tilde{J}_b + \tilde{J}_a \tilde{A}_b)\right]. \label{equation field equations}
\end{equation}
Therefore, the total energy-momentum tensor in the deformed manifold up to the second power of the electromagnetic 4-potential and the curvature quantities of the undeformed space-time is
\begin{equation}
\tilde{T}_{ab}=T_{ab}^{\rm{\langle M \rangle }}+\tilde{T}_{ab}^{\rm{\langle EM \rangle }}-(\tilde{A}_a \tilde{J}_b + \tilde{J}_a \tilde{A}_b).
\end{equation}
This energy-momentum tensor includes the energy-momenta of matter, the electromagnetic field, and additional terms involving the electric current. The latter show that there is an interaction between the electromagnetic potential and the electric current. 

Before seeing the implications of the field equations, we briefly discuss the signature of the model. If the 4D signature was taken to be ($-$ $+$ $+$ $+$), the sign of the electromagnetic energy-momentum tensor and inhomogeneous Maxwell's equations should be changed. Then, the resultant total energy-momentum tensor would not make sense since then it involves a difference between the energy-momentum of matter and that of the electromagnetic field. Therefore, ($+$ $-$ $-$ $-$) signature is correct for the 4D metric. This difference between two signatures originates from the deformation of the 5D manifold caused by the electromagnetic 4-potential. To see the effect of the signature in the 5D metric tensor, assume that the undeformed space-time is flat and the deformation is small enough so that index raising and lowering of the electromagnetic 4-potential can be done with a flat space-time metric. \(\tilde{A}^a = (\Phi, \vec{A})\) is common for both signature, so \(\tilde{A}_a = (\Phi, -\vec{A})\) for ($+$ $-$ $-$ $-$) and \(\tilde{A}_a = (-\Phi, \vec{A})\) for ($-$ $+$ $+$ $+$) (the tilde for \(\Phi\) and \(\vec{A}\) is omitted here). Then, the 5D deformed metric in each signature would be 
\begin{equation}
\tilde{g}_{\alpha \beta}=
\begin{pmatrix}
1+\Phi^2 & -\Phi \vec{A} & \Phi \\
-\Phi \vec{A} & -I + \vec{A} \otimes \vec{A} & -\vec{A} \\
\Phi & -\vec{A} & 1
\end{pmatrix} 
\quad \text{and} \quad
\tilde{g}_{\alpha \beta}=
\begin{pmatrix}
-1+\Phi^2 & -\Phi \vec{A} & -\Phi \\
-\Phi \vec{A} & I + \vec{A} \otimes \vec{A} & \vec{A} \\
-\Phi & \vec{A} & 1
\end{pmatrix}, 
\end{equation}
respectively. They differ not only in the signs but also in the absolute values of the 4D diagonal components. This gives the opposite sign for the electromagnetic energy-momentum in the total energy-momentum tensor. At a glance, time-like 5th dimension looks contrary to the discussion in \cite{bailin1987kaluza} that the 5th dimension should be space-like. However, this is not a contradiction but a result of different embedding. In Kaluza–Klein theory, the space-time metric is \(g_{ab}\) while the space-time metric in the present model is \(\tilde{g}_{ab}\), thus, the curvature quantities are calculated with different metrics. Because of this, while Kaluza–Klein theory observes that \(G_{ab}\) yields exact Einstein–Maxwell equations, the present model makes \(\bar{\tilde{G}}_{ab}\) to yield field equations containing Einstein–Maxwell equations. An interesting point in equation (\ref{equation Einstein tensor simplified}) is that the electromagnetic energy-momentum tensor should be transposed to the opposite side if one wants to get Einstein–Maxwell equations of \(G_{ab}\) by imposing the vacuum condition for \(\tilde{G}_{\alpha \beta}\). This implies different appropriate signs of electromagnetic energy-momentum tensor in field equations for two cases, and hence requires different signatures of the 5th dimension.

Now we go back to equation (\ref{equation field equations}). There are four points worth mentioning in the suggested field equations. First, the field equations relate a 4D part of a 5D tensor with a 4D tensor. This does not hold for general coordinate transformations in 5-dimensions but the field equations are justified by the covariance breaking discussed in section \ref{section covariance breaking}. Second, the field equations are 4D equations and its matter side (right-hand-side of equation (\ref{equation field equations})) is composed of the 4D energy-momentum tensor, not a 5D one. This is a consequence of the notion of particle-thread as discussed in section \ref{section construction}. Third, the field equations have new interaction terms compared to Einstein–Maxwell equations. Fourth, the suggested field equations become identical to Einstein's field equations in the electromagnetic field-free case.

\section{Unit system} \label{section unit system}
Before seeing the experimental implications of the model, we check our unit. In our system we take \(G=c=\mu_0/(16\pi)=16\pi \epsilon_0 = 1\). So, time, mass, and charge are all expressed in the length dimension. The conversion factors for mass and charge become
\begin{subequations}
\begin{eqnarray}
1 \ \text{kg}&=&\frac{\check{G}}{{\check c}^2}\ \text{m} \approx 7.42592 \times 10^{-28} \ \text{m}, \\
1 \ \text{C}&=&\sqrt{\frac{\check{\mu}_0\check{G}}{16\pi \check{c}^2}} \ \text{m} \approx 4.30869 \times 10^{-18} \ \text{m}, 
\end{eqnarray}
\end{subequations}
where the checked symbols refer to the numerical parts of the constants in the MKSC units. For electric potential, 
\begin{eqnarray}
1 \ \text{V} &=& \sqrt{\frac{16\pi \check{G} }{\check{\mu}_0 \check{c}^6}} \approx 1.91762 \times 10^{-27},
\label{naturalvolt}
\end{eqnarray}
which is dimensionless.

The Planck mass and the Planck charge become 
\begin{subequations}
\begin{eqnarray}
m_p&=&\sqrt{\frac{\check{G}\check{h}}{2\pi \check{c}^3}} \ \text{m} \approx 1.61623 \times 10^{-35}\ \text{m}, \\
q_p&=&\sqrt{\frac{\check{G}\check{h}}{8\pi \check{c}^3}}\ \text{m} \approx 8.08114 \times 10^{-36} \ \text{m}.
\end{eqnarray}
\end{subequations}
The Planck mass is twice the Planck charge.

\section{Experimental implications} \label{section experimental implications}
The effect of deformed space-time metric and the correction terms of the 4-acceleration are experimental observables for the test of the present model. For the explicit calculation of these two, the 4-potential should be known. As discussed in section \ref{section physical quantities}, the retarded electromagnetic 4-potential of the given electromagnetic source distribution in the undeformed space-time can be used as the zeroth order of the perturbation method under the assumption of that the 4-potential is small. This assumption is reasonable for the most cases since the conversion factor of the electric potential is order of \(10^{-27}\) as shown in section \ref{section unit system}. The following discussion will use particle picture in the space-time for the intuitive appeal.

For the simplicity, we take the electromagnetic source to be a spherically symmetric charge \(Q\) with mass \(M\) such that \(g_{ab}\) can be considered as a Schwarzschild metric. We assume that \(Q/r\) and \(M/r\) are both small parameters in the region of interest, so that the 4-potential becomes
\begin{equation}
\tilde{A}^a = 
\begin{pmatrix}
\frac{4Q}{r}, & 0, & 0, & 0
\end{pmatrix}
=\tilde{A}_a, 
\end{equation}
to the first order of those parameters. We proceed with this 4-potential for the following discussion. From now on, the tilde for the deformed space-time will be omitted in this section.

\subsection{Metric of the deformed space-time} \label{subsection metric of the deformed space-time}
We compare the metric of the deformed space-time against the Reissner–Nordström metric which is the metric of the space-time according to Einstein–Maxwell equations. Adopting the unit system of the present paper, the Reissner–Nordström metric is given by
\begin{equation}
ds^2=\left(1-\frac{2M}{r}+\frac{4Q^2}{r^2}\right)dt^2-\left(1-\frac{2M}{r}+\frac{4Q^2}{r^2}\right)^{-1}dr^2-r^2d\theta^2-r^2\sin^2(\theta)d\phi^2, 
\end{equation}
while the metric of the deformed space-time is
\begin{equation}
ds^2=\left(1-\frac{2M}{r}+\frac{16Q^2}{r^2}\right)dt^2-\left(1-\frac{2M}{r}\right)^{-1}dr^2-r^2d\theta^2-r^2\sin^2(\theta)d\phi^2.
\end{equation}
The effect of the charge becomes four times bigger in \(g_{00}\) and disappears in \(g_{11}\). This difference will give a chance to check which metric describes the phenomena better by measuring the bending of light ray passing near a charged object. 

The bending of light ray in the Reissner–Nordström metric is given at \cite{fernando2002gravitational} by using the method in \cite{weinberg1972gravitation}. The deflection angle \(\alpha\) is
\begin{equation}
\alpha = \frac{4M}{r_0}+\frac{4M^2}{r_0{}^2}\left(\frac{15\pi}{16}-1\right)-\frac{3\pi Q^2}{r_0{}^2}
\label{deflection angle RN}
\end{equation}
up to the second power of \(1/r_0\), where \(r_0\) denotes the closest approach distance of the light ray to the object \footnote[3]{The author of \cite{fernando2002gravitational} corrected the wrong term in the equation (12) in \cite{fernando2002gravitational} through a private communication.}. Here, we adopt the unit system of the present paper. By following the same procedure in \cite{fernando2002gravitational, weinberg1972gravitation}, the deflection angle for the metric of the deformed space-time becomes
\begin{equation}
\alpha = \frac{4M}{r_0}+\frac{4M^2}{r_0{}^2}\left(\frac{15\pi}{16}-1\right)-\frac{8\pi Q^2}{r_0{}^2}.
\label{deflection angle deformed spacetime}
\end{equation}
The effect of charge \(Q\) in the deflection angle is different in two cases by factor \(8/3\). The derivation of deflection angle is given in appendix \ref{appendix lightray}.

The equation between the impact parameter \(b\) and the closest approach distance \(r_0\) is given at \cite{weinberg1972gravitation} for a general metric. In the case of the Reissner–Nordström metric, it is
\begin{equation}
b=r_0 \left(1-\frac{2M}{r_0}+\frac{4Q^2}{r_0{}^2}\right)^{-1/2},
\end{equation}
and in the case of the deformed space-time metric, it is
\begin{equation}
b=r_0 \left(1-\frac{2M}{r_0}+\frac{16Q^2}{r_0{}^2}\right)^{-1/2}.
\end{equation}
If the measurable parameter is not \(r_0\) but \(b\), the above two equations can be used to obtain \(r_0\) for equations (\ref{deflection angle RN}) and (\ref{deflection angle deformed spacetime}).

\subsection{Correction terms in the 4-acceleration}
To calculate the explicit components of equation (\ref{equation geodesic projection final}), we consider a radial acceleration during a radial motion of a test particle with charge \(q\) and mass \(m\). The 4-velocity of the particle is 
\begin{equation}
U^a = 
\begin{pmatrix}
\gamma, & \gamma v, & 0, & 0
\end{pmatrix}, \quad \text{where} \quad \gamma =\left(1-\frac{2M}{r}+\frac{16Q^2}{r^2}-\frac{1}{1-\frac{2M}{r}}v^2\right)^{-1/2}.
\end{equation}
Now, for the electromagnetic field tensor, the only non-zero components of \(F^a_{\phantom{a}b}\) are
\begin{equation}
F^0_{\phantom{a}1}=\frac{4Q}{r^2}=F^1_{\phantom{a}0}
\end{equation}
so that the Lorentz force term of the 4-acceleration becomes
\begin{equation}
L^a = \frac{q}{m}\frac{4Q}{r^2}
\begin{pmatrix}
\gamma v, & \gamma, & 0, & 0
\end{pmatrix}. 
\end{equation}
Meanwhile, to the leading order, which is the lowest order in \(Q\), 
\begin{equation}
D^a = A^a - A_b\frac{dx^b}{d\tau}\frac{dx^a}{d\tau}=-\frac{4Q}{r}
\begin{pmatrix}
\gamma^2 v^2, & \gamma^2 v, & 0, & 0
\end{pmatrix}, 
\end{equation}
\begin{equation}
(\nabla_b A_c)\frac{dx^b}{d\tau}\frac{dx^c}{d\tau}=(\partial_1 A_0) \frac{dx^0}{d\tau}\frac{dx^1}{d\tau} = -\frac{4Q}{r^2}\gamma^2 v.
\end{equation}
Combining all these terms according to equation (\ref{equation geodesic projection final}), the 4-acceleration becomes
\begin{equation}
a^a=\frac{q}{m}\frac{4Q}{r^2}
\begin{pmatrix}
\gamma v, & \gamma, & 0, & 0
\end{pmatrix}
-\frac{q}{m}\frac{64Q^3}{r^4}\gamma^2 v^2
\begin{pmatrix}
\gamma v, & \gamma, & 0, & 0
\end{pmatrix}
+\frac{16Q^2}{r^3}\gamma^3 v^2
\begin{pmatrix}
\gamma v, & \gamma, & 0, & 0
\end{pmatrix}.
\end{equation}
The second and the third terms in the right-hand-side might give a chance to test the model experimentally.

\section{Conclusion and further research} \label{section conclusion}
In the present paper, the 5th dimension is introduced to explain the electromagnetism in terms of the geometric structure. The electromagnetic field-free 5D manifold is constructed by dragging the space-time. Then, electromagnetic 4-potential is interpreted as a shearing-like deformation factor of the 5D manifold and we obtain the resultant metric of the deformed 5D manifold which has the similar form of Kaluza–Klein metric with \(\phi^2=1\). But the embedment of the space-time is different and the metric employs a different electromagnetic potential.

The notion of the particle-thread naturally arises from the construction of the 5D manifold. A particle existing in the space-time will be dragged along the direction of the 5th axis as the 5D manifold is constructed from the dragging of the space-time. The particle-thread automatically guarantees the independence of the energy-momentum and the metric with respect to the 5th dimension. In addition to that, the particle-threads make it natural to use the 4D energy-momentum tensor along with the 4-current density for the description of 5D matter. By adopting the particle-thread, the electric charge of a particle is no longer a pre-given property but a quantity describing the motion of a particle-thread, that is to say the mass flow rate of a particle-thread along its extension. Furthermore, the mass flow to the 5th dimension results in the shearing of the 5D manifold to the 5th dimension. 

It is assumed that every portion of a particle-thread follows its 5D geodesic. Then a particle in the space-time is the cross section of the corresponding particle-thread along the space-time. A 4D trajectory of a particle in the space-time is the trace of the cross section of the particle-thread on the space-time which makes the 4D trajectory become the projection of a 5D geodesic onto the space-time. After the projection, its deviation from the space-time geodesic depends on the closeness between the direction of the 5D geodesic and the space-time. With this deviation, the projection in the deformed manifold results in the trajectory of a particle with the Lorentz force exerting on it. In addition to the Lorentz force term, there are additional small correction terms in the present model. The correction terms would be much smaller than the Lorentz force term unless the potential becomes extremely high. 

By examining the compatibility of reference frames it is concluded that the covariance of the 5th dimension is broken and there exists a common neutral plane. From the covariance breaking, there should be restrictions on the coordinate transformations in the 5-dimensions. These restrictions make the 4D part of a 5D tensor behaves as a 4D tensor. Since any relative motions in the space-time are parallel to the neutral plane, the covariance of the space-time still holds. The covariance breaking of the 5th dimension and the existence of the common neutral plane with the common 5th axis seem to originate from the dragging of the space-time.

The field equations between the 5D metric and the 5D matter are split into two equations, as a consequence of the particle-thread and the covariance breaking. One is Maxwell's equations in the deformed space-time, and the physically valid solution for these is the retarded potential in the deformed space-time. The other field equations that extend Einstein's field equations are suggested as \(\bar{\tilde{G}}_{ab}=8\pi \tilde{T}_{ab}\). The total energy-momentum tensor is the sum of energy-momenta of the matter, the electromagnetic field, and the interaction between the electromagnetic potential and the electric current. The last one is a newly introduced term. The 4D metric signature is fixed to be ($+$ $-$ $-$ $-$) to have the appropriate total energy-momentum tensor. This signature fixing makes the 5th dimension time-like.

Finally, the physical quantities for experimental verifications are suggested. One is the bending angle of a light ray and the other is the correction terms in the 4-acceleration. The calculations can be easily done by using the weak potential approximation.

The construction of this model leaves a few topics that require further research. First, the physical mechanism of dragging and a more fundamental cause of the covariance breaking should be discussed. Second, the reason why a specific directional degree of freedom is selected to be devoted for the dragging should be studied. One possibility is that the 5th dimension may have been evolved differently from the other four dimensions in the early universe. Third, there might exist a pre-given 5D energy-momentum even if it is improper under the construction of the 5D manifold and the particle-thread. Quantization of the 5th dimension may allow this by introducing a finite 5-volume density since then the thickness of the space-time along the 5th dimension becomes non-zero. Fourth, the possibility of a wave along the particle-thread is raised. If it exists, its property should be investigated. Lastly, numerical estimations of the correction terms are needed to compare with the experimental values.

\begin{acknowledgments}
TaeHun Kim gives special thanks to Hyunbyuk Kim, the corresponding author, for numerous communications and conversations which helped the research greatly. HbK also gave a number of detailed elaborations for the present paper. THK gives thanks to JinWoo Sung who were interested in the present topic and willing to have a discussion about it. JWS too gave detailed elaborations even in his military service. Finally, THK thanks for anonymous acquaintances who gave encouragements. 
\end{acknowledgments}

\appendix
\section{Notations} \label{appendix notations}
\begin{align*}
& \alpha, \ \beta, \ \gamma, \ ... = 0, \ 1, \ 2, \ 3, \ 4 && \text{Indices for 5D manifold} \nonumber\\
& a, \ b, \ c, \ ... = 0, \ 1, \ 2, \ 3 && \text{Indices for space-time} \nonumber\\
& d{x}^\alpha && \text{Infinitesimal displacement in both the deformed and the undeformed 5D manifold} \nonumber\\
& d{x}^a && \text{Infinitesimal displacement in both the deformed and the undeformed space-time} \nonumber\\
& \tilde{g}_{\alpha \beta} && \text{Metric of the deformed 5D manifold} \nonumber\\
& {g}_{\alpha \beta} && \text{Metric of the undeformed 5D manifold} \nonumber\\
& \tilde{g}_{ab} && \text{Metric of the deformed space-time} \nonumber\\
& {g}_{ab} && \text{Metric of the undeformed space-time} \nonumber\\
& \bar{\tilde{g}}_{ab} && \text{4D part of $\tilde{g}_{\alpha \beta}$} \nonumber\\
& \bar{\tilde{g}}_{ab} = \tilde{g}_{ab} && \text{The coordinate construction in the present paper} \nonumber\\
& \tilde{g}^{\alpha \beta} && \text{Inverse of $\tilde{g}_{\alpha \beta}$} \nonumber\\
& {g}^{\alpha \beta} && \text{Inverse of ${g}_{\alpha \beta}$} \nonumber\\
& \tilde{g}^{ab} && \text{Inverse of $\tilde{g}_{ab}$} \nonumber\\
& {g}^{ab} && \text{Inverse of ${g}_{ab}$} \nonumber\\
& \bar{\tilde{g}}^{ab} && \text{4D part of $\tilde{g}^{\alpha \beta}$} \nonumber\\
& d\tilde{s} && \text{Infinitesimal interval in the deformed 5D manifold} \nonumber\\
& d{s} && \text{Infinitesimal interval in the undeformed 5D manifold} \nonumber\\
& d\tilde{\tau} && \text{Infinitesimal proper time in the deformed space-time} \nonumber\\
& \tilde{\Gamma}^{\alpha}_{\phantom{\alpha} \beta \gamma} && \text{Christoffel symbols of the second kind of $\tilde{g}_{\alpha \beta}$} \nonumber\\
& \tilde{\Gamma}^{a}_{\phantom{a}bc} && \text{Christoffel symbols of the second kind of $\tilde{g}_{ab}$} \nonumber\\
& {\Gamma}^{a}_{\phantom{a}bc} && \text{Christoffel symbols of the second kind of ${g}_{ab}$} \nonumber\\
& \bar{\tilde{\Gamma}}^{a}_{\phantom{a}bc} && \text{4D part of $\tilde{\Gamma}^{\alpha}_{\phantom{\alpha} \beta \gamma}$} \nonumber\\
& \tilde{R}^{\alpha}_{\phantom{\alpha} \beta \gamma \delta} && \text{Riemann tensor of $\tilde{g}_{\alpha \beta}$} \nonumber\\
& \tilde{R}^{a}{}_{bcd} && \text{Riemann tensor of $\tilde{g}_{ab}$} \nonumber\\
& R^{a}_{\phantom{a}bcd} && \text{Riemann tensor of $g_{ab}$} \nonumber\\
& \bar{\tilde{R}}^{a}_{\phantom{a}bcd} && \text{4D part of $\tilde{R}^{\alpha}_{\phantom{\alpha} \beta \gamma \delta}$} \nonumber\\
& \tilde{R}_{\alpha \beta} && \text{Ricci tensor of $\tilde{R}^{\alpha}_{\phantom{\alpha} \beta \gamma \delta}$} \nonumber\\
& \tilde{R}_{ab} && \text{Ricci tensor of $\tilde{R}^{a}_{\phantom{a}bcd}$} \nonumber\\
& R_{ab} && \text{Ricci tensor of $R^{a}_{\phantom{a}bcd}$} \nonumber\\
& \bar{\tilde{R}}_{ab} && \text{4D part of $\tilde{R}_{\alpha \beta}$} \nonumber\\
& \bar{\tilde{R}} && \text{Ricci scalar of $\tilde{R}_{\alpha \beta}$} \nonumber\\
& \tilde{R} && \text{Ricci scalar of $\tilde{R}_{ab}$} \nonumber\\
& R && \text{Ricci scalar of $R_{ab}$} \nonumber\\
& \tilde{G}_{\alpha \beta} && \text{Einstein tensor of $\tilde{g}_{\alpha \beta}$, $\tilde{R}_{\alpha \beta}$, and $\bar{\tilde{R}}$} \nonumber\\
& \tilde{G}_{a b} && \text{Einstein tensor of $\tilde{g}_{a b}$, $\tilde{R}_{a b}$, and ${\tilde{R}}$ is never used in the present paper} \nonumber\\
& G_{ab} && \text{Einstein tensor of ${g}_{a b}$, ${R}_{a b}$, and $R$} \nonumber\\
& \bar{\tilde{G}}_{ab} && \text{4D part of $\tilde{G}_{\alpha \beta}$} \nonumber\\
\end{align*}\linebreak[0]

\section{Deformation} \label{appendix deformation}
While a metric deformation is considered in mathematical literature \cite{panda2012deformation, takashi1993deformation}, we introduce a formalism which can treat the change of manifold in a more direct and intuitive way. Before discussing our formalism, we give a brief remark about the metric deformation in each reference. In \cite{panda2012deformation}, the considered deformation is only about the metric, not about the manifold itself, that the metric of a flat 2D manifold undergoes a deformation by a coordinate transformation on the 2D manifold. In \cite{takashi1993deformation}, the deformation of a manifold itself is treated, but the considered deformation is to cut off a part of an original manifold and then replace it with some other same dimensional manifold possessing a different metric. These two are different from our formalism since the one used in this paper is a physical operation on a manifold itself, and it works as a continuous change of the manifold without cutting it off. 

A deformation in the present paper is a physical operation acting on a manifold such that the distance structure of the manifold, and hence the metric, changes keeping the previously established coordinate system. The idea can be illustrated through the following example. First, think of a physical 2D flat plane embedded in a 3D Euclidean space. In that plane, the assigned Cartesian coordinates can be visualized by orthogonal lines of constant \(x\) and constant \(y\). This plane can be deformed by an exerted stress on it, and can be turned into a curved surface, i.e. spherical or saddle surface. Then its intrinsic curvature will be changed, depending on the resultant surface. When the intrinsic curvature is changed, it cannot be a flat plane in \(\text{\textbf{R}}^3\) any more. For the curved resultant surface, we may put the original flat plane as a tangent at a chosen point. Near that point, to the first order, we can express the coordinates of the resultant surface in terms of the coordinates of the original plane. From the second order, however, this cannot be done since the non-vanishing intrinsic curvature depends on the second derivatives of the metric. This suggests that the first order derivatives of the resultant surface coordinates with respect to the original plane coordinates can be used to measure how much the resultant surface is \textit{deformed}. Also, these first order derivatives will not satisfy the commutativity of the second partial derivatives in general when the resultant surface is curved, because the resultant surface coordinates cannot be fully obtained through a coordinate transformation of the original plane coordinates.

Now extend the above idea to an N-dimensional manifold with a general metric. The embedding is no longer considered and only the intrinsic geometry is dealt with. If this manifold is deformed, the deformation could be expressed by quantities resembling the first-order derivatives, as in the previous 2D example. This is the deformation operator which is expressed as \(D^{\gamma}_{\alpha}\) in equation (\ref{general deformation}). Then how can the effect of the deformation be expressed in terms of the quantities associated with the manifold? Recalling the 2D example, the first-order derivatives at a point were obtained when the deformed and the undeformed ones become tangent at that point. In this setup, they share a common tangent space at the point and the metric of the deformed surface at that point will be obtained via the usual tensor transformation rule with the transformation matrix composed of these first-order derivatives. Therefore, for general cases, it is reasonable to deduce equation (\ref{general deformation}) that resembles a coordinate transformation to express how the deformation affects the metric, again, because the first-order derivatives of the coordinates represent the deformation. But then, the crucial difference between the deformation operator and the coordinate transformation matrix is their derivatives. There is no condition for the derivative of a deformation operator. But for the coordinate transformation matrix, there is a condition forced to satisfy which is the commutativity of the second partial derivatives between the new and the old coordinates. In other words, a deformation resembles a coordinate transformation in such a way that for each single point of a manifold its effect can be treated as a coordinate transformation, even though the coordinate transformation which is valid for the whole deformed manifold cannot be achieved while a deformation affects the whole manifold. This is why we cannot obtain the Riemann tensor, the Ricci tensor, and the Ricci scalar of the deformed manifold by regarding the deformation operator as a coordinate transformation matrix. Those quantities involve derivatives of the metric tensor which require not only one point but also the nearby points of the manifold.

The deformation is a physical operation on a manifold, so actually it would be better to be called a manifold deformation, but since the effect of a deformation is always expressed in terms of a metric, we use the term deformation only. And generically, the deformation of a manifold does not rely on a higher dimensional ambient manifold. In the present model, a deformation of the 5D manifold is caused internally by flowing particle-threads in the 5D manifold. 

To explicitly see that the deformation is a physical operation on a manifold, we consider a deformation applied to a 2D flat plane which resembles a coordinate transformation rescaling each axis, or in other words, a deformation that stretches or compresses the plane along the direction of each axis. This kind of deformation, with index \(\alpha\) and \(\beta\) running from 1 to 2, will be expressed as 
\begin{equation}
D_\alpha^\beta=
\begin{cases}
f(x, y), & \text{if} \ \alpha=\beta=1 \\
g(x, y) & \text{if} \ \alpha=\beta=2 \\
0, & \text{otherwise}
\end{cases}
\end{equation}
so that the metric of the deformed 2D surface to be 
\begin{equation}
d\tilde{s}^2 = f^2(x, y)dx^2+g^2(x, y)dy^2.
\end{equation}
For this metric, the Gaussian curvature is
\begin{equation}
K=\frac{f^2\left(f_{,2}g_{,2}-f_{,22}g\right)+g^2\left(f_{,1}g_{,1}-fg_{,11}\right)}{f^3g^3}, 
\end{equation}
which is not necessarily vanishing and its value depends on the details of the deformation unlike the vanishing Gaussian curvature of the original 2D flat plane. 

Another example of the stretching and compressing deformation in the space-time is the Schwarzschild metric. The Schwarzschild metric with mass \(M\) has the form of 
\begin{equation}
ds^2 = \left(1-\frac{2M}{r}\right)dt^2-\frac{1}{1-\frac{2M}{r}}dr^2-r^2d\theta^2-r^2\sin^2\theta d\phi^2.
\end{equation}
The Riemann tensor of Schwarzschild metric does not vanish so the metric cannot be a result of a coordinate transformation from the flat space-time metric. However, the Schwarzschild metric and the flat space-time metric have an identical angular part in a spherical coordinate and both are diagonalized. Therefore, the deformation which resembles the coordinate transformation of rescaling the time and the radial coordinate can make the Schwarzschild metric from the flat space-time metric. In this case, the deformation is done by
\begin{equation}
D_a^b=
\begin{cases}
\left(1-\frac{2M}{r}\right)^{\frac{1}{2}}, & \text{if} \ a=b=0 \\
\left(1-\frac{2M}{r}\right)^{-\frac{1}{2}} & \text{if} \ a=b=1 \\
1 & \text{if} \ a=b=2,\ 3\\
0, & \text{otherwise}.
\end{cases}
\end{equation}
If we force interpretation of \(D_{a}^{b}\) as a coordinate transformation matrix and denote the `Schwarzschild coordinates' as primed ones and the `flat coordinates' as unprimed ones, it is readily checkable that the second partial derivatives of \(t'\) with respect to \(t\) and \(r\) do not commute. Hence \(D_{a}^{b}\) cannot be a coordinate transformation matrix, but represents a deformation.

\section{\(f(\tilde{\tau})\)} \label{appendix f tau}
Computations (except equation (\ref{f tilde tau final})) in this appendix are not limited to the metric used in the present paper. They are able to be applied to any other metric. We expect these computations will be useful for future reference.

It is well known that equation (\ref{equation geodesic tau}) requires
\begin{eqnarray}
f(\tilde{\tau})=\frac{d^2 \tilde{s} / d\tilde{\tau}^2}{d\tilde{s} / d\tilde{\tau}}.
\end{eqnarray}
Now, using 
\begin{eqnarray}
\frac{d\tilde{s}}{d\tilde{\tau}}= \sqrt{\tilde{g}_{\alpha \beta} \frac{dx^{\alpha}}{d\tilde{\tau}} \frac{dx^{\beta}}{d\tilde{\tau}}}
\end{eqnarray}
gives 
\begin{eqnarray}
f(\tilde{\tau})=\frac{\frac{d}{d\tilde{\tau}}\left(\tilde{g}_{\alpha \beta} \frac{dx^{\alpha}}{d\tilde{\tau}} \frac{dx^{\beta}}{d\tilde{\tau}}\right)}{2\tilde{g}_{\alpha \beta} \frac{dx^{\alpha}}{d\tilde{\tau}} \frac{dx^{\beta}}{d\tilde{\tau}}}.
\end{eqnarray}
Since \(\bar{\tilde{g}}_{ab}=\tilde{g}_{ab}\) and \(d\tilde{\tau}^2=\tilde{g}_{ab}dx^{a}dx^{b}\), 
\begin{equation}
\tilde{g}_{\alpha \beta} \frac{dx^{\alpha}}{d\tilde{\tau}} \frac{dx^{\beta}}{d\tilde{\tau}}=1+2\tilde{g}_{a4}\frac{dx^{a}}{d\tilde{\tau}}\frac{dx^{4}}{d\tilde{\tau}}+\tilde{g}_{44}\frac{dx^{4}}{d\tilde{\tau}}\frac{dx^{4}}{d\tilde{\tau}}.
\label{gdxdxdtdt}
\end{equation}
Therefore, splitting the numerator gives
\begin{eqnarray}
f(\tilde{\tau})=\frac{\frac{d}{d\tilde{\tau}}\left(2\tilde{g}_{a4}\frac{dx^{a}}{d\tilde{\tau}}\frac{dx^{4}}{d\tilde{\tau}}+\tilde{g}_{44}\frac{dx^{4}}{d\tilde{\tau}}\frac{dx^{4}}{d\tilde{\tau}}\right)}{2\tilde{g}_{\alpha \beta} \frac{dx^{\alpha}}{d\tilde{\tau}} \frac{dx^{\beta}}{d\tilde{\tau}}}.
\end{eqnarray}
By using the Leibniz rule of differentiation, 
\begin{widetext}
\begin{eqnarray}
f(\tilde{\tau})=
\frac{1}{2\tilde{g}_{\alpha \beta}\frac{dx^{\alpha}}{d\tilde{\tau}} \frac{dx^{\beta}}{d\tilde{\tau}}}
\left(2\tilde{g}_{a4,\gamma}\frac{dx^{a}}{d\tilde{\tau}}\frac{dx^{4}}{d\tilde{\tau}}\frac{dx^{\gamma}}{d\tilde{\tau}}
+2\tilde{g}_{a4}\frac{d^2 x^{a}}{d\tilde{\tau}^2}\frac{dx^{4}}{d\tilde{\tau}}
+2\tilde{g}_{a4}\frac{dx^{a}}{d\tilde{\tau}}\frac{d^2 x^{4}}{d\tilde{\tau}^2}
+\tilde{g}_{44,\gamma}\frac{dx^{4}}{d\tilde{\tau}}\frac{dx^{4}}{d\tilde{\tau}}\frac{dx^{\gamma}}{d\tilde{\tau}}
+2\tilde{g}_{44}\frac{dx^{4}}{d\tilde{\tau}}\frac{d^2 x^{4}}{d\tilde{\tau}^2}\right). \nonumber \\
\end{eqnarray}
There are three second derivatives of \(x\) with respect to \(\tilde{\tau}\). We apply equation (\ref{equation geodesic tau}) to them. The result is

\begin{eqnarray}
f(\tilde{\tau})=
\frac{1}{2\tilde{g}_{\alpha \beta}\frac{dx^{\alpha}}{d\tilde{\tau}} \frac{dx^{\beta}}{d\tilde{\tau}}}&&
\left[
2\tilde{g}_{a4,\gamma}\frac{dx^{a}}{d\tilde{\tau}}\frac{dx^{4}}{d\tilde{\tau}}\frac{dx^{\gamma}}{d\tilde{\tau}}
+2\tilde{g}_{a4}\left(f(\tilde{\tau})\frac{dx^{a}}{d\tilde{\tau}}-\tilde{\Gamma}^{a}_{\phantom{a} \beta \gamma}\frac{dx^{\beta}}{d\tilde{\tau}}\frac{dx^{\gamma}}{d\tilde{\tau}}\right)\frac{dx^{4}}{d\tilde{\tau}} \right. \nonumber \\ &&\hphantom{[}+ \left. 2\tilde{g}_{a4}\frac{dx^{a}}{d\tilde{\tau}}\left(f(\tilde{\tau})\frac{dx^{4}}{d\tilde{\tau}}-\tilde{\Gamma}^{4}_{\phantom{4} \beta \gamma}\frac{dx^{\beta}}{d\tilde{\tau}}\frac{dx^{\gamma}}{d\tilde{\tau}}\right)
+\tilde{g}_{44,\gamma}\frac{dx^{4}}{d\tilde{\tau}}\frac{dx^{4}}{d\tilde{\tau}}\frac{dx^{\gamma}}{d\tilde{\tau}} \right. \nonumber \\ &&\hphantom{[}+\left. 
2\tilde{g}_{44}\frac{dx^{4}}{d\tilde{\tau}}\left(f(\tilde{\tau})\frac{dx^{4}}{d\tilde{\tau}}-\tilde{\Gamma}^{4}_{\phantom{4} \beta \gamma}\frac{dx^{\beta}}{d\tilde{\tau}}\frac{dx^{\gamma}}{d\tilde{\tau}}\right)
\right],
\end{eqnarray}\\
and transposing all of the \(f(\tilde{\tau})\)s in the right-hand-side to the left-hand-side gives

\begin{eqnarray}
&&\left(
\tilde{g}_{\alpha \beta} \frac{dx^{\alpha}}{d\tilde{\tau}}\frac{dx^{\beta}}{d\tilde{\tau}}
-2\tilde{g}_{a4}\frac{dx^{a}}{d\tilde{\tau}}\frac{dx^{4}}{d\tilde{\tau}}
-\tilde{g}_{44}\frac{dx^{4}}{d\tilde{\tau}}\frac{dx^{4}}{d\tilde{\tau}} \right)
f(\tilde{\tau})
\nonumber \\
&&=
\tilde{g}_{a4,\gamma}\frac{dx^{a}}{d\tilde{\tau}}\frac{dx^{4}}{d\tilde{\tau}}\frac{dx^{\gamma}}{d\tilde{\tau}}
-\tilde{g}_{a4}\tilde{\Gamma}^{a}_{\phantom{a} \beta \gamma}\frac{dx^{\beta}}{d\tilde{\tau}}\frac{dx^{\gamma}}{d\tilde{\tau}}\frac{dx^{4}}{d\tilde{\tau}}
-\tilde{g}_{a4}\tilde{\Gamma}^{4}_{\phantom{4} \beta \gamma}\frac{dx^{\beta}}{d\tilde{\tau}}\frac{dx^{\gamma}}{d\tilde{\tau}}\frac{dx^{a}}{d\tilde{\tau}}
+\frac{1}{2}\tilde{g}_{44,\gamma}\frac{dx^{4}}{d\tilde{\tau}}\frac{dx^{4}}{d\tilde{\tau}}\frac{dx^{\gamma}}{d\tilde{\tau}}
-\tilde{g}_{44}\tilde{\Gamma}^{4}_{\phantom{4} \beta \gamma}\frac{dx^{\beta}}{d\tilde{\tau}}\frac{dx^{\gamma}}{d\tilde{\tau}}\frac{dx^{4}}{d\tilde{\tau}}. \nonumber \\
\end{eqnarray}
However, the left-hand-side is just equal to \(f(\tilde{\tau})\) due to equation (\ref{gdxdxdtdt}). Therefore,  
\begin{eqnarray}
f(\tilde{\tau})
=
\tilde{g}_{a4,\gamma}\frac{dx^{a}}{d\tilde{\tau}}\frac{dx^{4}}{d\tilde{\tau}}\frac{dx^{\gamma}}{d\tilde{\tau}}
-\tilde{g}_{a4}\tilde{\Gamma}^{a}_{\phantom{a} \beta \gamma}\frac{dx^{\beta}}{d\tilde{\tau}}\frac{dx^{\gamma}}{d\tilde{\tau}}\frac{dx^{4}}{d\tilde{\tau}}
-\tilde{g}_{a4}\tilde{\Gamma}^{4}_{\phantom{4} \beta \gamma}\frac{dx^{\beta}}{d\tilde{\tau}}\frac{dx^{\gamma}}{d\tilde{\tau}}\frac{dx^{a}}{d\tilde{\tau}}
+\frac{1}{2}\tilde{g}_{44,\gamma}\frac{dx^{4}}{d\tilde{\tau}}\frac{dx^{4}}{d\tilde{\tau}}\frac{dx^{\gamma}}{d\tilde{\tau}}
-\tilde{g}_{44}\tilde{\Gamma}^{4}_{\phantom{4} \beta \gamma}\frac{dx^{\beta}}{d\tilde{\tau}}\frac{dx^{\gamma}}{d\tilde{\tau}}\frac{dx^{4}}{d\tilde{\tau}}. \nonumber \\
\end{eqnarray}
The second and the last term can be combined to give a term with the Christoffel symbols of the first kind:
\begin{equation}
f(\tilde{\tau})
=
\tilde{g}_{a4,\gamma}\frac{dx^{a}}{d\tilde{\tau}}\frac{dx^{4}}{d\tilde{\tau}}\frac{dx^{\gamma}}{d\tilde{\tau}}
-\tilde{\Gamma}_{4 \beta \gamma}\frac{dx^{\beta}}{d\tilde{\tau}}\frac{dx^{\gamma}}{d\tilde{\tau}}\frac{dx^{4}}{d\tilde{\tau}}
-\tilde{g}_{a4}\tilde{\Gamma}^{4}_{\phantom{4} \beta \gamma}\frac{dx^{\beta}}{d\tilde{\tau}}\frac{dx^{\gamma}}{d\tilde{\tau}}\frac{dx^{a}}{d\tilde{\tau}}
+\frac{1}{2}\tilde{g}_{44,\gamma}\frac{dx^{4}}{d\tilde{\tau}}\frac{dx^{4}}{d\tilde{\tau}}\frac{dx^{\gamma}}{d\tilde{\tau}}.
\end{equation}
Until here, the details of \(\tilde{g}_{\alpha \beta}\) are never used. To obtain a simple expression in the case of the metric of the deformed 5D manifold, expand the Christoffel symbols of the first kind in terms of the metric. After expanding them, eliminate the terms involving \(\tilde{g}_{\alpha \beta ,4}\) and \(\tilde{g}_{44,\alpha}\) which are zero. As a result, the first and the second terms cancel each other, and the last term vanishes. Finally,
\begin{equation}
f(\tilde{\tau})
=
-\tilde{g}_{a4}\tilde{\Gamma}^{4}_{\phantom{4} \beta \gamma}\frac{dx^{\beta}}{d\tilde{\tau}}\frac{dx^{\gamma}}{d\tilde{\tau}}\frac{dx^{a}}{d\tilde{\tau}}.
\label{f tilde tau final}
\end{equation}

\section{The Christoffel symbols} \label{appendix Christoffel deformed}
From the metric and the inverse metric of the deformed 5D manifold given at section \ref{section physical quantities}, the Christoffel symbols of the deformed 5D manifold can be expressed in terms of the Christoffel symbols of the deformed space-time and electromagnetic 4-potential.

\begin{subequations}
\begin{eqnarray}
\bar{\tilde{\Gamma}}^{a}_{\phantom{a}bc}&=&\frac{1}{2}\tilde{g}^{a\delta}(\tilde{g}_{\delta b,c}+\tilde{g}_{\delta c,b}-\tilde{g}_{cb,\delta}) \nonumber \\ &=&\frac{1}{2}\bar{\tilde{g}}^{ad}(\tilde{g}_{db,c}+\tilde{g}_{dc,b}-\tilde{g}_{cb,d})+\frac{1}{2}\tilde{g}^{a4}(\tilde{g}_{4b,c}+\tilde{g}_{4c,b}-\tilde{g}_{cb,4}) \nonumber \\ &=& \tilde{\Gamma}^a_{\phantom{a}bc}+\frac{1}{1-\tilde{A}_e\tilde{A}^e}\tilde{A}^a\tilde{A}^d\tilde{\Gamma}_{dbc}-\frac{1}{2}\frac{1}{1-\tilde{A}_e \tilde{A}^e}\tilde{A}^a(\tilde{A}_{b,c}+\tilde{A}_{c,b}) \nonumber \\ &=& \tilde{\Gamma}^a_{\phantom{a}bc}-\frac{1}{2}\frac{1}{1-\tilde{A}_e \tilde{A}^e}\tilde{A}^a(\tilde{\nabla}_c \tilde{A}_{b}+\tilde{\nabla}_b \tilde{A}_{c}),\\
\tilde{\Gamma}^a_{\phantom{a}4b}&=&\frac{1}{2}\tilde{g}^{a\delta}(\tilde{g}_{\delta 4,b}+\tilde{g}_{\delta b,4}-\tilde{g}_{b4,\delta})=\frac{1}{2}\bar{\tilde{g}}^{ad}(\tilde{g}_{d4,b}-\tilde{g}_{b4,d}) \nonumber \\ &=& -\frac{1}{2}\left(\tilde{F}^a_{\phantom{a}b}+\frac{1}{1-\tilde{A}_e \tilde{A}^e}\tilde{A}^a \tilde{A}^d \tilde{F}_{db}\right), \\
\tilde{\Gamma}^\alpha_{\phantom{a}44}&=&\frac{1}{2}\tilde{g}^{\alpha \delta}(\tilde{g}_{\delta 4,4}+\tilde{g}_{\delta 4,4}-\tilde{g}_{44,\delta})=0, \\
\tilde{\Gamma}^4_{\phantom{a}bc}&=&\frac{1}{2}\tilde{g}^{4\delta}(\tilde{g}_{\delta b,c}+\tilde{g}_{\delta c,b}-\tilde{g}_{bc,\delta}) \nonumber \\ &=& \frac{1}{2}\tilde{g}^{4d}(\tilde{g}_{db,c}+\tilde{g}_{dc,b}-\tilde{g}_{cb,d})+\frac{1}{2}\tilde{g}^{44}(\tilde{g}_{4b,c}+\tilde{g}_{4c,b}-\tilde{g}_{cb,4}) \nonumber \\ &=& -\frac{1}{1-\tilde{A}_e \tilde{A}^e}\tilde{A}^d\tilde{\Gamma}_{dbc}+\frac{1}{2}\frac{1}{1-\tilde{A}_e \tilde{A}^e}(\tilde{A}_{b,c}+\tilde{A}_{c,b}) \nonumber \\ &=& \frac{1}{2}\frac{1}{1-\tilde{A}_e \tilde{A}^e}(\tilde{\nabla}_c \tilde{A}_b + \tilde{\nabla}_b \tilde{A}_c), \\ 
\tilde{\Gamma}^4_{\phantom{a}4b}&=&\frac{1}{2}\tilde{g}^{4\delta}(\tilde{g}_{\delta 4,b}+\tilde{g}_{\delta b,4}-\tilde{g}_{b4,\delta})=\frac{1}{2}\tilde{g}^{4d}(\tilde{g}_{d4,b}-\tilde{g}_{b4,d}) \nonumber \\ &=& \frac{1}{2}\frac{1}{1-\tilde{A}_e \tilde{A}^e}\tilde{A}^d\tilde{F}_{db}, 
\end{eqnarray}
\end{subequations}
where
\begin{equation}
\tilde{F}_{ab} = \tilde{\nabla}_a\tilde{A}_b - \tilde{\nabla}_b \tilde{A}_a = \tilde{A}_{b,a}-\tilde{A}_{a,b}.
\end{equation}

For the relation between the Christoffel symbols of the undeformed space-time (\(\Gamma^a_{\phantom{a}bc}\)) and those of the deformed space-time (\(\tilde{\Gamma}^a_{\phantom{a}bc}\)), we need to express the metric and the inverse metric of the undeformed space-time in terms of those of the deformed space-time:
\begin{subequations}
\begin{eqnarray}
g_{ab}&=&\tilde{g}_{ab}-\tilde{A}_a\tilde{A}_b, \\
g^{ab}&=&\tilde{g}^{ab}+\frac{1}{1-\tilde{A}_e \tilde{A}^e}\tilde{A}^a \tilde{A}^b,
\end{eqnarray}
\end{subequations}
where the second equation is obtained by applying the result of \cite{miller1981inverse} to the first equation. Since the deformation does not change any coordinate values, the ordinary derivatives of \(g_{ab}\) are identical in both undeformed space-time and deformed space-time. Having this in mind, the equation between \(\Gamma^a_{\phantom{a}bc}\) and \(\tilde{\Gamma}^a_{\phantom{a}bc}\) becomes
\begin{eqnarray}
\Gamma^a_{\phantom{a}bc}&=&\frac{1}{2}g^{ad}(g_{db,c}+g_{dc,b}-g_{bc,d}) \nonumber \\
&=&\frac{1}{2}\left(\tilde{g}^{ad}+\frac{1}{1-\tilde{A}_e \tilde{A}^e}\tilde{A}^a\tilde{A}^d\right)(\tilde{g}_{db,c}+\tilde{g}_{dc,b}-\tilde{g}_{bc,d}-\tilde{A}_{d,c}\tilde{A}_{b}-\tilde{A}_{d}\tilde{A}_{b,c}-\tilde{A}_{d,b}\tilde{A}_{c}-\tilde{A}_d\tilde{A}_{c,b}+\tilde{A}_{b,d}\tilde{A}_c+\tilde{A}_b\tilde{A}_{c,d}) \nonumber \\
&=&\tilde{\Gamma}^a_{\phantom{a}bc}+\frac{1}{1-\tilde{A}_e \tilde{A}^e}\tilde{A}^a\tilde{A}^d\tilde{\Gamma}_{dbc}+\frac{1}{2}(\tilde{A}_{b}\tilde{F}^a_{\phantom{a}c}+\tilde{A}_{c}\tilde{F}^a_{\phantom{a}b})-\frac{1}{2}\tilde{A}^a(\tilde{A}_{b,c}+\tilde{A}_{c,b}) \nonumber \\ &&+\frac{1}{2}\frac{1}{1-\tilde{A}_e \tilde{A}^e}\tilde{A}^a\tilde{A}^d(\tilde{A}_b\tilde{F}_{dc}+\tilde{A}_c\tilde{F}_{db}-\tilde{A}_d\tilde{A}_{b,c}-\tilde{A}_d\tilde{A}_{c,b}) \nonumber \\
&=&\tilde{\Gamma}^a_{\phantom{a}bc}+\frac{1}{2}(\tilde{A}_{b}\tilde{F}^a_{\phantom{a}c}+\tilde{A}_{c}\tilde{F}^a_{\phantom{a}b})-\frac{1}{2}\tilde{A}^a(\tilde{\nabla}_c\tilde{A}_{b}+\tilde{\nabla}_b\tilde{A}_{c})+\frac{1}{2}\frac{1}{1-\tilde{A}_e \tilde{A}^e}\tilde{A}^a\tilde{A}^d(\tilde{A}_b\tilde{F}_{dc}+\tilde{A}_c\tilde{F}_{db}-\tilde{A}_d\tilde{\nabla}_c\tilde{A}_{b}-\tilde{A}_d\tilde{\nabla}_b\tilde{A}_{c}) \nonumber \\
&=&\tilde{\Gamma}^a_{\phantom{a}bc}+\frac{1}{2}(\tilde{A}_{b}\tilde{F}^a_{\phantom{a}c}+\tilde{A}_{c}\tilde{F}^a_{\phantom{a}b})-\frac{1}{2}\frac{1}{1-\tilde{A}_e \tilde{A}^e}\tilde{A}^a(\tilde{\nabla}_c\tilde{A}_{b}+\tilde{\nabla}_b\tilde{A}_{c})+\frac{1}{2}\frac{1}{1-\tilde{A}_e \tilde{A}^e}\tilde{A}^a\tilde{A}^d(\tilde{A}_b\tilde{F}_{dc}+\tilde{A}_c\tilde{F}_{db}). \label{space-time christoffel}
\end{eqnarray}

Using the above equation, we can also write \(\bar{\tilde{\Gamma}}^a_{\phantom{a}bc}\) in terms of \(\Gamma^a_{\phantom{a}bc}\), which is
\begin{equation}
\bar{\tilde{\Gamma}}^a_{\phantom{a}bc}=\Gamma^a_{\phantom{a}bc}-\frac{1}{2}(\tilde{A}_{b}\tilde{F}^a_{\phantom{a}c}+\tilde{A}_{c}\tilde{F}^a_{\phantom{a}b})-\frac{1}{2}\frac{1}{1-\tilde{A}_e \tilde{A}^e}\tilde{A}^a\tilde{A}^d(\tilde{A}_b\tilde{F}_{dc}+\tilde{A}_c\tilde{F}_{db}).
\end{equation}

\section{Space-time embedment check via Ricci scalar} \label{appendix Ricci scalar}
The Ricci scalar for each metric \(g_{ab}\) and \(\tilde{g}_{ab} = g_{ab}+\tilde{A}_{a}\tilde{A}_{b}\) can be computed by using equation (\ref{space-time christoffel}) in appendix \ref{appendix Christoffel deformed}. If \(R \neq \tilde{R}\), the two metrics cannot express the same 4D hypersurface, so the embedding of the space-time in the present model and in Kaluza–Klein theory will be different. The calculation is done up to the second power of \(\tilde{A}_a\) and higher order terms are excluded in the following equations. Starting from equation (\ref{space-time christoffel}), the Riemann tensor becomes
\begin{eqnarray}
&&R^a{}_{bcd} \nonumber \\
&&= \Gamma^a{}_{bd,c}-\Gamma^a{}_{bc,d}+\Gamma^a{}_{ce}\Gamma^e{}_{bd}-\Gamma^a{}_{de}\Gamma^e{}_{bc} \nonumber \\
&&= \tilde{\Gamma}^a{}_{bd,c}+\frac{1}{2}\left(\tilde{A}_{b,c}\tilde{F}^a{}_{d}+\tilde{A}_b\tilde{F}^a{}_{d,c}+\tilde{A}_{d,c}\tilde{F}^a{}_{b}+\tilde{A}_{d}\tilde{F}^a{}_{b,c}\right)-\frac{1}{2}\tilde{A}^a{}_{,c}\left(\tilde{\nabla}_{d}\tilde{A}_{b}+\tilde{\nabla}_b \tilde{A}_d\right)-\frac{1}{2}\tilde{A}^a\left(\tilde{\nabla}_d\tilde{A}_{b}+\tilde{\nabla}_b\tilde{A}_d\right)_{,c} \nonumber \\
&& \phantom{=}-\tilde{\Gamma}^a{}_{bc,d}-\frac{1}{2}\left(\tilde{A}_{b,d}\tilde{F}^a{}_{c}+\tilde{A}_b\tilde{F}^a{}_{c,d}+\tilde{A}_{c,d}\tilde{F}^a{}_{b}+\tilde{A}_{c}\tilde{F}^a{}_{b,d}\right)+\frac{1}{2}\tilde{A}^a{}_{,d}\left(\tilde{\nabla}_{c}\tilde{A}_{b}+\tilde{\nabla}_b \tilde{A}_c\right)+\frac{1}{2}\tilde{A}^a\left(\tilde{\nabla}_c\tilde{A}_{b}+\tilde{\nabla}_b\tilde{A}_c\right)_{,d} \nonumber \\
&& \phantom{=} +\tilde{\Gamma}^a{}_{ce}\tilde{\Gamma}^e{}_{bd}+\frac{1}{2}\tilde{\Gamma}^a{}_{ce}\left[\tilde{A}_b\tilde{F}^e{}_d+\tilde{A}_d\tilde{F}^e{}_b-\tilde{A}^e\left(\tilde{\nabla}_b\tilde{A}_d+\tilde{\nabla}_d\tilde{A}_{b}\right)\right]+\frac{1}{2}\tilde{\Gamma}^e{}_{bd}\left[\tilde{A}_c\tilde{F}^a{}_e+\tilde{A}_e\tilde{F}^a{}_c-\tilde{A}^a\left(\tilde{\nabla}_c\tilde{A}_e+\tilde{\nabla}_e\tilde{A}_c\right)\right] \nonumber \\
&& \phantom{=} -\tilde{\Gamma}^a{}_{de}\tilde{\Gamma}^e{}_{bc}-\frac{1}{2}\tilde{\Gamma}^a{}_{de}\left[\tilde{A}_b\tilde{F}^e{}_c+\tilde{A}_c\tilde{F}^e{}_b-\tilde{A}^e\left(\tilde{\nabla}_b\tilde{A}_c+\tilde{\nabla}_c\tilde{A}_{b}\right)\right]-\frac{1}{2}\tilde{\Gamma}^e{}_{bc}\left[\tilde{A}_d\tilde{F}^a{}_e+\tilde{A}_e\tilde{F}^a{}_d-\tilde{A}^a\left(\tilde{\nabla}_d\tilde{A}_e+\tilde{\nabla}_e\tilde{A}_d\right)\right] \nonumber \\
&&= \tilde{R}^a{}_{bcd}+\frac{1}{2}\left(\tilde{F}_{cd}\tilde{F}^a{}_b+\tilde{A}_b\tilde{\nabla}_c\tilde{F}^a{}_d-\tilde{A}_b\tilde{\nabla}_d\tilde{F}^a{}_c+\tilde{F}^a{}_d\tilde{\nabla}_c\tilde{A}_b-\tilde{F}^a{}_c\tilde{\nabla}_d\tilde{A}_b+\tilde{A}_d\tilde{\nabla}_c\tilde{F}^a{}_b-\tilde{A}_c\tilde{\nabla}_d\tilde{F}^a{}_b\right) \nonumber \\
&& \phantom{=} -\frac{1}{2}\left(\tilde{\nabla}_c\tilde{A}^a\right)\left(\tilde{\nabla}_d\tilde{A}_b+\tilde{\nabla}_b\tilde{A}_d\right)+\frac{1}{2}\left(\tilde{\nabla}_d\tilde{A}^a\right)\left(\tilde{\nabla}_c\tilde{A}_b+\tilde{\nabla}_b\tilde{A}_c\right) \nonumber \\
&& \phantom{=} -\frac{1}{2}\tilde{A}^a\tilde{\nabla}_c\left(\tilde{\nabla}_d\tilde{A}_b+\tilde{\nabla}_b\tilde{A}_d\right)+\frac{1}{2}\tilde{A}^a\tilde{\nabla}_d\left(\tilde{\nabla}_c\tilde{A}_b+\tilde{\nabla}_b\tilde{A}_c\right).
\end{eqnarray}
By remembering that \(\tilde{A}_a\) is the retarded solution in the Lorenz gauge for Maxwell's equations in the deformed space-time, we have \(\tilde{\nabla}_a\tilde{A}^a=0\). Using this, the Ricci tensor becomes
\begin{eqnarray}
R_{bd} &=& \tilde{R}_{bd}+\frac{1}{2}\left(\tilde{F}_{ad}\tilde{F}^a{}_b+\tilde{A}_b\tilde{\nabla}_a\tilde{F}^a{}_d+\tilde{A}_d\tilde{\nabla}_a\tilde{F}^a{}_b\right)+\frac{1}{2}\left(\tilde{\nabla}_a\tilde{A}_b\right)\left(\tilde{\nabla}^a\tilde{A}_d-\tilde{\nabla}_d\tilde{A}^a\right)-\frac{1}{2}\tilde{A}^a\tilde{\nabla}_d\left(\tilde{\nabla}_a\tilde{A}_b-\tilde{\nabla}_b\tilde{A}_a\right) \nonumber \\
&&+\frac{1}{2}\left(\tilde{\nabla}_d\tilde{A}^a\right)\left(\tilde{\nabla}_a\tilde{A}_b+\tilde{\nabla}_b\tilde{A}_a\right)-\frac{1}{2}\tilde{A}^a\tilde{\nabla}_a\left(\tilde{\nabla}_d\tilde{A}_b+\tilde{\nabla}_b\tilde{A}_d\right)+\frac{1}{2}\tilde{A}^a\tilde{\nabla}_d\left(\tilde{\nabla}_a\tilde{A}_b+\tilde{\nabla}_b\tilde{A}_a\right) \nonumber \\
&=& \tilde{R}_{bd}+\frac{1}{2}\left(\tilde{F}_{ad}\tilde{F}^a{}_b+\tilde{A}_b\tilde{\nabla}_a\tilde{F}^a{}_d+\tilde{A}_d\tilde{\nabla}_a\tilde{F}^a{}_b\right)+\frac{1}{2}\left(\tilde{\nabla}_a\tilde{A}_b\right)\left(\tilde{\nabla}^a\tilde{A}_d\right)+\frac{1}{2}\left(\tilde{\nabla}_d\tilde{A}^a\right)\left(\tilde{\nabla}_b\tilde{A}_a\right)\nonumber \\
&&-\frac{1}{2}\tilde{A}^a\tilde{\nabla}_a\left(\tilde{\nabla}_d\tilde{A}_b+\tilde{\nabla}_b\tilde{A}_d\right)+\tilde{A}^a\tilde{\nabla}_d\tilde{\nabla}_b\tilde{A}_a.
\end{eqnarray}
Note that the last term is symmetric for the indices \(d\) and \(b\) by the anti-symmetric property of the Riemann tensor. Now, for the Ricci scalar, 
\begin{eqnarray}
R&=&R_{bd}g^{bd}=R_{bd}\left(\tilde{g}^{bd}+\frac{1}{1-\tilde{A}_e\tilde{A}^e}\tilde{A}^b\tilde{A}^d\right) \nonumber \\
&=&\tilde{R}+\frac{1}{2}\tilde{F}_{ab}\tilde{F}^{ab}+\tilde{A}_b\left(\tilde{\nabla}_a\tilde{F}^{ab}\right)+\left(\tilde{\nabla}_a\tilde{A}_b\right)\left(\tilde{\nabla}^a\tilde{A}^b\right)+\tilde{A}^a\tilde{\nabla}^b\tilde{\nabla}_b\tilde{A}_a+\tilde{R}_{ab}\tilde{A}^a\tilde{A}^b \nonumber \\
&=&\tilde{R}+\frac{1}{2}\tilde{F}_{ab}\tilde{F}^{ab}+\left(\tilde{\nabla}_a\tilde{A}_b\right)\left(\tilde{\nabla}^a\tilde{A}^b\right)+2\tilde{A}^a\tilde{\nabla}^b\tilde{\nabla}_b\tilde{A}_a \nonumber \\
&\neq&\tilde{R}.
\end{eqnarray}
Thus, the embedment of the space-time into the 5D manifold is different for two metrics  \(g_{ab}\) and \(\tilde{g}_{ab} = g_{ab}+\tilde{A}_{a}\tilde{A}_{b}\).

\section{4-acceleration} \label{appendix geodesic projection}
Making equation (\ref{equation geodesic projection initial}) to be equation (\ref{equation geodesic projection final}) is done by using the Christoffel symbols in appendix \ref{appendix Christoffel deformed}. From equation (\ref{equation geodesic projection initial}), 
\begin{eqnarray}
&&\frac{d^2x^a}{d\tilde{\tau}^2}+\left[\tilde{\Gamma}^a_{\phantom{a}bc}-\frac{1}{2}\frac{1}{1-\tilde{A}_e \tilde{A}^e}\tilde{A}^a(\tilde{\nabla}_c\tilde{A}_b+\tilde{\nabla}_b\tilde{A}_c)\right]\frac{dx^b}{d\tilde{\tau}}\frac{dx^c}{d\tilde{\tau}} \nonumber \\ &&=-2\tilde{\Gamma}^a_{\phantom{a}4b}\frac{dx^4}{d\tilde{\tau}}\frac{dx^b}{d\tilde{\tau}}-\tilde{\Gamma}^a_{\phantom{a}44}\frac{dx^4}{d\tilde{\tau}}\frac{dx^4}{d\tilde{\tau}} \nonumber \\ &&\phantom{=}-\left(\tilde{g}_{4d}\tilde{\Gamma}^4_{\phantom{a}bc}\frac{dx^b}{d\tilde{\tau}}\frac{dx^c}{d\tilde{\tau}}\frac{dx^d}{d\tilde{\tau}}\right)\frac{dx^a}{d\tilde{\tau}}-2\left(\tilde{g}_{4d}\tilde{\Gamma}^4_{\phantom{a}4b}\frac{dx^4}{d\tilde{\tau}}\frac{dx^b}{d\tilde{\tau}}\frac{dx^d}{d\tilde{\tau}}\right)\frac{dx^a}{d\tilde{\tau}}-\left(\tilde{g}_{4d}\tilde{\Gamma}^4_{\phantom{a}44}\frac{dx^4}{d\tilde{\tau}}\frac{dx^4}{d\tilde{\tau}}\frac{dx^d}{d\tilde{\tau}}\right)\frac{dx^a}{d\tilde{\tau}}.
\end{eqnarray}
Expanding and rearranging all the Christoffel symbols in the right-hand-side using the results of appendix \ref{appendix Christoffel deformed} give
\begin{eqnarray}
\tilde{a}^a&=&\tilde{F}^a_{\phantom{a}b}\frac{dx^4}{d\tilde{\tau}}\frac{dx^b}{d\tilde{\tau}}+\frac{1}{1-\tilde{A}_e \tilde{A}^e}\tilde{A}^a\tilde{A}^d\tilde{F}_{db}\frac{dx^4}{d\tilde{\tau}}\frac{dx^b}{d\tilde{\tau}}+\frac{1}{2}\frac{1}{1-\tilde{A}_e \tilde{A}^e}\tilde{A}^a(\tilde{\nabla}_c\tilde{A}_b+\tilde{\nabla}_b\tilde{A}_c)\frac{dx^b}{d\tilde{\tau}}\frac{dx^c}{d\tilde{\tau}} \nonumber \\ && -\frac{1}{2}\frac{1}{1-\tilde{A}_e \tilde{A}^e}(\tilde{\nabla}_c\tilde{A}_b+\tilde{\nabla}_b\tilde{A}_c)\frac{dx^b}{d\tilde{\tau}}\frac{dx^c}{d\tilde{\tau}}\tilde{A}_d\frac{dx^d}{d\tilde{\tau}}\frac{dx^a}{d\tilde{\tau}}-\frac{1}{1-\tilde{A}_e \tilde{A}^e}\tilde{A}^f\tilde{F}_{fb}\frac{dx^4}{d\tilde{\tau}}\frac{dx^b}{d\tilde{\tau}}\tilde{A}_d\frac{dx^d}{d\tilde{\tau}}\frac{dx^a}{d\tilde{\tau}} \nonumber \\ 
&=&\tilde{L}^a+\frac{1}{1-\tilde{A}_e \tilde{A}^e}\tilde{A}_d\tilde{L}^d\tilde{A}^a+\frac{1}{1-\tilde{A}_e \tilde{A}^e}(\tilde{\nabla}_b\tilde{A}_c)\frac{dx^b}{d\tilde{\tau}}\frac{dx^c}{d\tilde{\tau}}\tilde{A}^a \nonumber \\ && -\frac{1}{1-\tilde{A}_e \tilde{A}^e}(\tilde{\nabla}_b\tilde{A}_c)\frac{dx^b}{d\tilde{\tau}}\frac{dx^c}{d\tilde{\tau}}\tilde{A}_d\frac{dx^d}{d\tilde{\tau}}\frac{dx^a}{d\tilde{\tau}}-\frac{1}{1-\tilde{A}_e \tilde{A}^e}\tilde{A}_f\tilde{L}^f\tilde{A}_d\frac{dx^d}{d\tilde{\tau}}\frac{dx^a}{d\tilde{\tau}} \nonumber \\ 
&=&\tilde{L}^a+\frac{1}{1-\tilde{A}_e \tilde{A}^e}\tilde{L}^c\tilde{A}_c\tilde{D}^a+\frac{1}{1-\tilde{A}_e \tilde{A}^e}(\tilde{\nabla}_b\tilde{A}_c)\frac{dx^b}{d\tilde{\tau}}\frac{dx^c}{d\tilde{\tau}}\tilde{D}^a,
\end{eqnarray}
where $\tilde{a}^a$, $\tilde{L}^a$, and $\tilde{D}^a$ are defined in equation (\ref{aLD in 4-acc}).

\section{The Einstein tensor} \label{appendix Einstein tensor}
The 5D metric in \cite{kerner2000geodesic} is the metric of the deformed 5D manifold of the present paper while the 4D metric in \cite{kerner2000geodesic} is the metric of the undeformed space-time. Also the electromagnetic 4-potential is defined on the undeformed space-time in \cite{kerner2000geodesic}. For this case, the Christoffel symbols of the deformed 5D manifold in \cite{kerner2000geodesic}, with adapted notations and orders, are, 
\begin{subequations}
\begin{eqnarray}
\bar{\tilde{\Gamma}}^{a}_{\phantom{a}bc}&=&\Gamma^{a}_{\phantom{a}bc}-\frac{1}{2}(A_c F^{a}_{\phantom{a}b} + A_b F^{a}_{\phantom{a}c}),\\
\tilde{\Gamma}^{a}_{\phantom{a}4b}&=&-\frac{1}{2}F^{a}_{\phantom{a} b},\\
\tilde{\Gamma}^{\alpha}_{\phantom{\alpha}44}&=&0, \\
\tilde{\Gamma}^{4}_{\phantom{4}bc}&=&\frac{1}{2}(\nabla_b A_c + \nabla_c A_b)-\frac{1}{2}A^d(A_c F_{bd} + A_b F_{cd}), \\
\tilde{\Gamma}^{4}_{\phantom{4}4b}&=&-\frac{1}{2}A^d F_{bd}.
\end{eqnarray}
\end{subequations}
From these, the Riemann tensors in \cite{kerner2000geodesic} are, 
\begin{subequations}
\begin{eqnarray} 
\bar{\tilde{R}}^{a}_{\phantom{a}bcd}&=&R^{a}_{\phantom{a}bcd}+\frac{1}{4}(F_{c}^{\phantom{c}a} F_{bd}-F_{d}^{\phantom{d}a} F_{bc} +2F_{b}^{\phantom{b}a} F_{cd})-\frac{1}{2}A_b \nabla^a F_{cd}+\frac{1}{2}(A_d \nabla_c F_{b}^{\phantom{b}a} - A_c \nabla_d F_{b}^{\phantom{b}a}) \nonumber \\ &&+\frac{1}{4}A_b F_{e}^{\phantom{e}a}(A_c F_{d}^{\phantom{d}e}-A_d F_{c}^{\phantom{c}e}), \\
\tilde{R}^{4}_{\phantom{4}a4b}&=&\frac{1}{4}F_{a}^{\phantom{a}c} F_{bc} +\frac{1}{2}A_c \nabla_b F_{a}^{\phantom{a}c} -\frac{1}{4}A_a A^c F_{b}^{\phantom{b}e} F_{ec}, \\
\tilde{R}^{a}_{\phantom{a}4bc}&=&-\frac{1}{2}\nabla^a F_{bc} +\frac{1}{4}F_{e}^{\phantom{e}a}(A_b F_{c}^{\phantom{c}e} - A_c F_{b}^{\phantom{b}e}), \\
\tilde{R}^{4}_{\phantom{4}44a}&=&-\frac{1}{4}A_d F_{e}^{\phantom{e}d} F_{a}^{\phantom{a}e}, \\
\tilde{R}^{a}_{\phantom{a}4b4}&=&-\frac{1}{4}F_{e}^{\phantom{e}a} F_{b}^{\phantom{b}e}.
\end{eqnarray}
\end{subequations}

However, in our case, the electromagnetic 4-potential \(\tilde{A}_a\) is defined on the deformed space-time. Thus the Riemann tensors should be expressed in terms of \(\tilde{A}_a\) and the covariant derivatives \(\tilde{\nabla}_a\) should also be defined in the deformed space-time. Fortunately, the Christoffel symbols obtained in appendix \ref{appendix Christoffel deformed} and in \cite{kerner2000geodesic} differ only by the tilde mark on the electromagnetic 4-potential and disagree from the third order of the 4-potential. It can be checked by a direct computation that we can use those results up to the second order of the 4-potential and the Riemann tensor of the deformed space-time by replacing all the \(A_a\) s, \(F_{ab}\) s, and \(\nabla_a\) s with the \(\tilde{A}_a\) s, \(\tilde{F}_{ab}\) s, and \(\tilde{\nabla}_a\) s. Again, following computations are all valid up to the second order of the 4-potential and the curvature quantities of the deformed space-time.

The Ricci tensor becomes,
\begin{subequations}
\begin{eqnarray}
\bar{\tilde{R}}_{ab}&=&\tilde{R}^{\alpha}_{\phantom{\alpha} a \alpha b}=\bar{\tilde{R}}^{c}_{\phantom{c}acb}+\tilde{R}^{4}_{\phantom{4}a4b} \nonumber \\ 
&=& R^{c}_{\phantom{c}acb}+\frac{1}{4}(- \tilde{F}_{b}^{\phantom{b}c} \tilde{F}_{ac} + 2\tilde{F}_{a}^{\phantom{a}c} \tilde{F}_{cb})-\frac{1}{2}\tilde{A}_a \tilde{\nabla}^c \tilde{F}_{cb} +\frac{1}{2}(\tilde{A}_b \tilde{\nabla}_c \tilde{F}_{a}^{\phantom{a}c} - \tilde{A}_c \tilde{\nabla}_b \tilde{F}_{a}^{\phantom{a}c}) + \frac{1}{4}\tilde{F}_{a}^{\phantom{a}c} \tilde{F}_{bc} +\frac{1}{2}\tilde{A}_c \tilde{\nabla}_b \tilde{F}_{a}^{\phantom{a}c} \nonumber \\ 
&=& R_{ab} - \frac{1}{2}\tilde{F}_{a}^{\phantom{a}c} \tilde{F}_{bc} -\frac{1}{2}(\tilde{A}_a \tilde{\nabla}_c \tilde{F}^{c}_{\phantom{c}b} + \tilde{A}_b \tilde{\nabla}_c \tilde{F}^{c}_{\phantom{c}a}),   \\
\tilde{R}_{4a}&=&\tilde{R}^{\alpha}_{\phantom{\alpha} 4 \alpha a}=\tilde{R}^{b}_{\phantom{b}4ba}+\tilde{R}^{4}_{\phantom{4}44a} \nonumber \\
&=&-\frac{1}{2}\tilde{\nabla}^b \tilde{F}_{ba}, \\
\tilde{R}_{44}&=&\tilde{R}^{\alpha}_{\phantom{\alpha} 4 \alpha 4}=\tilde{R}^{a}_{\phantom{a}4a4}+\tilde{R}^{4}_{\phantom{4}444}=\tilde{R}^{a}_{\phantom{a}4a4} \nonumber \\
&=&\frac{1}{4}\tilde{F}_{ea}\tilde{F}^{ea}.
\end{eqnarray}
\end{subequations}
Next, the Ricci scalar becomes, 
\begin{eqnarray}
\bar{\tilde{R}}&=&\tilde{g}^{\alpha\beta}\tilde{R}_{\alpha\beta} =\bar{\tilde{g}}^{ab}\bar{\tilde{R}}_{ab} +2\tilde{g}^{a4}\tilde{R}_{a4}+\tilde{g}^{44}\tilde{R}_{44}  \nonumber \\
&=&\bar{\tilde{g}}^{ab}\left[R_{ab} - \frac{1}{2}\tilde{F}_{a}^{\phantom{a}c} \tilde{F}_{bc} -\frac{1}{2}\left(\tilde{A}_a \tilde{\nabla}_c \tilde{F}^{c}_{\phantom{c}b} + \tilde{A}_b \tilde{\nabla}_c \tilde{F}^{c}_{\phantom{c}a}\right)\right] +2\tilde{g}^{a4}\left(-\frac{1}{2}\tilde{\nabla}^c \tilde{F}_{ca}\right)+\tilde{g}^{44}\left(\frac{1}{4}\tilde{F}_{ec}\tilde{F}^{ec}\right) \nonumber \\
&=&R-\frac{1}{2}\tilde{F}_{ab}\tilde{F}^{ab}-\tilde{A}_a\tilde{\nabla}_c \tilde{F}^{ca} + \tilde{A}^a \tilde{\nabla}^c \tilde{F}_{ca}+ \frac{1}{4} \tilde{F}_{ec} \tilde{F}^{ec}\nonumber \\
&=&R-\frac{1}{4}\tilde{F}_{ab}\tilde{F}^{ab}.
\end{eqnarray}
Finally, the 4D part of the Einstein tensor of the deformed 5D manifold can be obtained:
\begin{eqnarray}
\bar{\tilde{G}}_{ab}&=&\bar{\tilde{R}}_{ab}-\frac{1}{2}\bar{\tilde{g}}_{ab}\bar{\tilde{R}}  \nonumber \\ 
&=&R_{ab} - \frac{1}{2}\tilde{F}_{a}^{\phantom{a}c} \tilde{F}_{bc} -\frac{1}{2}(\tilde{A}_a \tilde{\nabla}_c \tilde{F}^{c}_{\phantom{c}b} + \tilde{A}_b \tilde{\nabla}_c \tilde{F}^{c}_{\phantom{c}a})-\frac{1}{2}(g_{ab}+\tilde{A}_a \tilde{A}_b)\left(R-\frac{1}{4}\tilde{F}_{ec}\tilde{F}^{ec}\right)  \nonumber \\
&=&G_{ab} +\frac{1}{2}\left(\frac{1}{4}\tilde{g}_{ab}\tilde{F}_{ec}\tilde{F}^{ec}-\tilde{F}_{a}^{\phantom{a}c} \tilde{F}_{bc}\right)-\frac{1}{2}(\tilde{A}_a \tilde{\nabla}_c \tilde{F}^{c}_{\phantom{c}b} + \tilde{A}_b \tilde{\nabla}_c \tilde{F}^{c}_{\phantom{c}a}).
\end{eqnarray}
\end{widetext}

\section{Light ray bending} \label{appendix lightray}
From the work of \cite{weinberg1972gravitation}, the deflection angle \(\alpha\) for light ray passing near the origin of the metric
\begin{equation}
ds^2 = B(r)dt^2-A(r)dr^2-r^2d\theta^2-r^2\sin^2(\theta)d\phi^2
\end{equation}
is given by
\begin{equation}
\alpha=2\left\{\int_{r_0}^\infty A^{1/2}(r)\left[\left(\frac{r}{r_0}\right)^2\left(\frac{B(r_0)}{B(r)}\right)-1\right]^{-1/2}\frac{dr}{r}\right\}-\pi,
\end{equation}
where \(r_0\) is the minimum value of \(r\) within the path of the light ray. We want to expand \(\alpha\) in power series of \(1/r_0\). For the two cases in section \ref{subsection metric of the deformed space-time}, the \(r_0\) appears in the form of \(M/r_0\) or \(Q/r_0\), so we can expand the integrand in power series of \(M\) and \(Q\). To see the effect of charge in its leading order, the series expansion is made up to the second order. 

For the Reissner–Nordström metric, 
\begin{subequations}
\begin{eqnarray}
A(r)&=&\left(1-\frac{2M}{r}+\frac{4Q^2}{r^2}\right)^{-1}, \\
B(r)&=&1-\frac{2M}{r}+\frac{4Q^2}{r^2}, 
\end{eqnarray}
\end{subequations}
so, 
\begin{eqnarray}
\frac{B(r_0)}{B(r)}
&=&\left(1-\frac{2M}{r_0}+\frac{4Q^2}{r_0{}^2}\right)\left(1-\frac{2M}{r}+\frac{4Q^2}{r^2}\right)^{-1} \nonumber \\
&=&\left(1-\frac{2M}{r_0}+\frac{4Q^2}{r_0{}^2}\right)\left(1+\frac{2M}{r}+\frac{4M^2}{r^2}-\frac{4Q^2}{r^2}+O(X^3)\right) \nonumber \\
&=&1+2M\left(\frac{1}{r}-\frac{1}{r_0}\right)+4M^2\frac{1}{r}\left(\frac{1}{r}-\frac{1}{r_0}\right)+4Q^2\left(\frac{1}{r_0{}^2}-\frac{1}{r^2}\right)+O(X^3), 
\end{eqnarray}
where \(X\) denotes \(M\) or \(Q\). This gives
\begin{eqnarray}
\left(\frac{r}{r_0}\right)^2\left(\frac{B(r_0)}{B(r)}\right)-1
&=&\left(\frac{r}{r_0}\right)^2\left[1+2M\left(\frac{1}{r}-\frac{1}{r_0}\right)+\frac{4M^2}{r}\left(\frac{1}{r}-\frac{1}{r_0}\right)+4Q^2\left(\frac{1}{r_0{}^2}-\frac{1}{r^2}\right)+O(X^3)\right]-1 \nonumber \\
&=&\left[\left(\frac{r}{r_0}\right)^2-1\right]\left[1-\frac{2Mr}{r_0(r+r_0)}-\frac{4M^2}{r_0(r+r_0)}+\frac{4Q^2}{r_0{}^2}+O(X^3)\right],
\end{eqnarray}
so the integral becomes
\begin{eqnarray}
&&\int_{r_0}^\infty A^{1/2}(r)\left[\left(\frac{r}{r_0}\right)^2\left(\frac{B(r_0)}{B(r)}\right)-1\right]^{-1/2}\frac{dr}{r} \nonumber \\
&&=\int_{r_0}^\infty \frac{dr}{r\left[\left(\frac{r}{r_0}\right)^2-1\right]^{1/2}} \left(1-\frac{2M}{r}+\frac{4Q^2}{r^2}\right)^{-1/2}\left[1-\frac{2Mr}{r_0(r+r_0)}-\frac{4M^2}{r_0(r+r_0)}+\frac{4Q^2}{r_0{}^2}+O(X^3)\right]^{-1/2} \nonumber \\
&&=\int_{r_0}^\infty \frac{dr}{r\left[\left(\frac{r}{r_0}\right)^2-1\right]^{1/2}} \left(1+\frac{M}{r}+\frac{3M^2}{2r^2}-\frac{2Q^2}{r^2}+O(X^3)\right) \nonumber \\
&&\phantom{=\int_{r_0}^\infty \frac{dr}{r\left[\left(\frac{r}{r_0}\right)^2-1\right]^{1/2}}}\times \left[1+\frac{Mr}{r_0(r+r_0)}+\frac{2M^2}{r_0(r+r_0)}+\frac{3M^2r^2}{2r_0{}^2(r+r_0)^2}-\frac{2Q^2}{r_0{}^2}+O(X^3)\right] \nonumber \\
&&=\int_{r_0}^\infty \frac{dr}{r\left[\left(\frac{r}{r_0}\right)^2-1\right]^{1/2}}\left[1+\frac{M}{r}+\frac{Mr}{r_0(r+r_0)}+\frac{3M^2}{2r^2}+\frac{3M^2}{r_0(r+r_0)}+\frac{3M^2r^2}{2r_0{}^2(r+r_0)^2}-\frac{2Q^2}{r^2}-\frac{2Q^2}{r_0{}^2}+O(X^3)\right] \nonumber \\
&&=\frac{\pi}{2}+\frac{2M}{r_0}+\frac{2M^2}{r_0{}^2}\left(\frac{15\pi}{16}-1\right)-\frac{3\pi Q^2}{2r_0{}^2} + O(X^3)
\end{eqnarray}
which gives the deflection angle
\begin{equation}
\alpha=\frac{4M}{r_0}+\frac{4M^2}{r_0{}^2}\left(\frac{15\pi}{16}-1\right)-\frac{3\pi Q^2}{r_0{}^2}
\end{equation}
up to the second order.

For the deformed space-time metric, 
\begin{subequations}
\begin{eqnarray}
A(r)&=&\left(1-\frac{2M}{r}\right)^{-1}, \\
B(r)&=&1-\frac{2M}{r}+\frac{16Q^2}{r^2}, 
\end{eqnarray}
\end{subequations}
so from the previous calculation, we can immediately obtain 
\begin{equation}
\left(\frac{r}{r_0}\right)^2\left(\frac{B(r_0)}{B(r)}\right)-1
=\left[\left(\frac{r}{r_0}\right)^2-1\right]\left[1-\frac{2Mr}{r_0(r+r_0)}-\frac{4M^2}{r_0(r+r_0)}+\frac{16Q^2}{r_0{}^2}+O(X^3)\right]
\end{equation}
and the integral becomes
\begin{eqnarray}
&&\int_{r_0}^\infty A^{1/2}(r)\left[\left(\frac{r}{r_0}\right)^2\left(\frac{B(r_0)}{B(r)}\right)-1\right]^{-1/2}\frac{dr}{r} \nonumber \\
&&=\int_{r_0}^\infty \frac{dr}{r\left[\left(\frac{r}{r_0}\right)^2-1\right]^{1/2}} \left(1-\frac{2M}{r}\right)^{-1/2}\left[1-\frac{2Mr}{r_0(r+r_0)}-\frac{4M^2}{r_0(r+r_0)}+\frac{16Q^2}{r_0{}^2}+O(X^3)\right]^{-1/2} \nonumber \\
&&=\int_{r_0}^\infty \frac{dr}{r\left[\left(\frac{r}{r_0}\right)^2-1\right]^{1/2}} \left(1+\frac{M}{r}+\frac{3M^2}{2r^2}+O(X^3)\right) \nonumber \\
&&\phantom{=\int_{r_0}^\infty \frac{dr}{r\left[\left(\frac{r}{r_0}\right)^2-1\right]^{1/2}}}\times \left[1+\frac{Mr}{r_0(r+r_0)}+\frac{2M^2}{r_0(r+r_0)}+\frac{3M^2r^2}{2r_0{}^2(r+r_0)^2}-\frac{8Q^2}{r_0{}^2}+O(X^3)\right] \nonumber \\
&&=\int_{r_0}^\infty \frac{dr}{r\left[\left(\frac{r}{r_0}\right)^2-1\right]^{1/2}}\left[1+\frac{M}{r}+\frac{Mr}{r_0(r+r_0)}+\frac{3M^2}{2r^2}+\frac{3M^2}{r_0(r+r_0)}+\frac{3M^2r^2}{2r_0{}^2(r+r_0)^2}-\frac{8Q^2}{r_0{}^2}+O(X^3)\right] \nonumber \\
&&=\frac{\pi}{2}+\frac{2M}{r_0}+\frac{2M^2}{r_0{}^2}\left(\frac{15\pi}{16}-1\right)-\frac{4\pi Q^2}{r_0{}^2} +O(X^3)
\end{eqnarray}
which gives the deflection angle
\begin{equation}
\alpha=\frac{4M}{r_0}+\frac{4M^2}{r_0{}^2}\left(\frac{15\pi}{16}-1\right)-\frac{8\pi Q^2}{r_0{}^2}
\end{equation}
up to the second order as well.

\bibliography{ArticleNotes}
\end{document}